\documentstyle[aas2pp4,psfig]{article}
\newcommand{\solarmass}{\mbox{${\rm M_{\odot}}$}}

\def\ltsima{$\; \buildrel < \over \sim \;$}
\def\simlt{\lower.5ex\hbox{\ltsima}}
\def\gtsima{$\; \buildrel > \over \sim \;$}
\def\simgt{\lower.5ex\hbox{\gtsima}}

\newcommand{\grs}{{GRS 1915+105}}
\newcommand{\etal}{{et al.}}
\newcommand{\Kp}{{$K^{\prime}$}}
\newcommand{\Ks}{{$Ks$}}
\newcommand{\J}{{$J$}}
\renewcommand{\H}{{$H$}}
\newcommand{\K}{{$K$}}
\newcommand{\ergs}{ergs cm$^{-2}$ s$^{-1}$}

\newcommand{\asca}{{\it ASCA}}
\newcommand{\rxte}{{\it RXTE}}
\newcommand{\sax}{{\it BeppoSAX}}
\newcommand{\cgro}{{\it CGRO}}
\newcommand{\chandra}{{\it Chandra}}

\lefthead{Ueda et al.}
\righthead{Multiwavelength Campaign of \grs\ }

\begin{document}

\title{Study of the Largest Multiwavelength Campaign of the 
Microquasar \grs }

\author{Y.~Ueda\altaffilmark{1},
K.~Yamaoka\altaffilmark{2},
C.~S\'anchez-Fern\'andez\altaffilmark{3},
V.~Dhawan\altaffilmark{4},
S.~Chaty\altaffilmark{5},
J.E.~Grove\altaffilmark{6},
M.~McCollough\altaffilmark{7},
A.J.~Castro-Tirado\altaffilmark{3,8},
F.~Mirabel\altaffilmark{9,10},
K.~Kohno\altaffilmark{11,12},
M.~Feroci\altaffilmark{13},
P.~Casella\altaffilmark{13},
S.A.~Trushkin\altaffilmark{14},
H.~Castaneda\altaffilmark{15},
J.~Rodr\'{\i}guez\altaffilmark{9},
P.~Durouchoux\altaffilmark{9},
K.~Ebisawa\altaffilmark{16},
T.~Kotani\altaffilmark{16},
J.~Swank\altaffilmark{16},
H.~Inoue\altaffilmark{1}
}

\altaffiltext{1}{Institute of Space and Astronautical Science, Sagamihara 229-8510, Japan}
\altaffiltext{2}{Institute of Physical and Chemical Research (RIKEN), Wako 351-01, Japan}
\altaffiltext{3}{LAEFF-INTA, Madrid, Spain}
\altaffiltext{4}{NRAO, USA}
\altaffiltext{5}{Department of Physics and Astronomy, The Open University, UK}
\altaffiltext{6}{Naval Research Lab, USA}
\altaffiltext{7}{MSFC/NASA, USA}
\altaffiltext{8}{IAA-CSIC, Granada, Spain}
\altaffiltext{9}{Service d'Astrophysique, Saclay, France}
\altaffiltext{10}{IAFE/CONICET, Argentina}
\altaffiltext{11}{Nobeyama Radio Observatory, NAO, Japan}
\altaffiltext{12}{Institute of Astronomy, The University of Tokyo, Japan}
\altaffiltext{13}{Istituto di Astrofisica Spaziale, C.N.R., Italy}
\altaffiltext{14}{SAO, Russia}
\altaffiltext{15}{Observatorio Astronomico Nacional de Mexico, Mexico}
\altaffiltext{16}{GSFC/NASA, USA}

\newpage

\begin{abstract}

We present the results from a multiwavelength campaign of \grs\
performed from 2000 April 16 to 25. This is one of the largest
coordinated set of observations ever performed for this source,
covering the wide energy band in radio (13.3--0.3 cm), near-infrared
($J$-$H$-$K$), X-rays and Gamma-rays (from 1 keV to 10 MeV).  During
the campaign \grs\ was predominantly in the ``plateau'' (or low/hard)
state but sometimes showed soft X-ray oscillations: before April
20.3, rapid, quasi-periodic ($\simeq$45 min) flare-dip cycles were
observed.
In the spectral energy distribution in the plateau state,
optically-thick synchrotron emission and Comptonization is dominant in
the radio and X- to Gamma-ray bands, respectively. The small
luminosity in the radio band relative to that in X-rays indicates that
\grs\ falls in the regime of ``radio-quiet'' microquasars.
In three epochs we detected faint flares in the radio or infrared
bands with amplitudes of 10--20 mJy. The radio flares observed on
April 17 shows frequency-dependent peak delay, consistent with an
expansion of synchrotron-emitting region starting at the transition
from the hard-dip to the soft-flare states in X-rays. On the other
hand, infrared flares on April 20 appear to follow (or precede) the
beginning of X-ray oscillations with an inconstant time delay of
$\simeq$ 5--30 min. This implies that the infrared emitting region
is located far from the black hole by $\simgt 10^{13}$ cm, while
its size is  $\simlt 10^{12}$ cm constrained from the time variability. 
We find a
good correlation between the quasi-steady flux level in the
near-infrared band and in the X-ray band. From this we estimate that the
reprocessing of X-rays, probably occurring in the outer parts of the accretion
disk, accounts for about 20--30\% of the observed \K\ 
magnitude in the plateau state.
The OSSE spectrum in the 0.05--10 MeV band is represented by a single
power law with a photon index of $3.1$ extending to $\sim$1 MeV with
no cutoff. We can model the combined GIS-PCA-HEXTE spectra covering
1--200 keV by a sum of the multi-color disk model, a broken power law
modified with a high-energy cutoff, and a reflection component. The
power-law slope above $\sim$30 keV is found to be very similar between
different states in spite of large flux variations in soft X-rays,
implying that the electron energy distribution is not affected by the
change of the state in the accretion disk.

\end{abstract}


\section{Introduction}

Up to present a few superluminal jet sources have been identified in
our Galaxy (for a review see Mirabel \& Rodr\'{\i}guez 1999): \grs ,
GRO~J1655--40, possibly XTE J1748--288, XTE J1550--564 (Hannikainen
\etal\ 2001), and V4641~Sgr (Orosz \etal\ 2001). The similarity of the
observed properties in these sources to those of AGNs indicates that
the underlying physics is common to a wide mass range of the central
black hole from $\sim 10 \solarmass$ to $\sim 10^{9} \solarmass$,
except for scaling in luminosity, size, and variability, which are
proportional to the black hole mass. These objects, so-called
microquasars, are ideal laboratories for understanding the mechanism
of relativistic jets and its relation with accretion flow onto a black
hole, owing to the brightness in X-ray bands and fast
variability. Studies in multiwavelength bands are particularly
important, because with X-rays and Gamma-rays we can probe into the
innermost part of the accretion disk, while in radio and infrared
bands we can trace ejection of relativistic jets.

\grs\ was discovered in 1992 as an X-ray transient by WATCH / {\it
GRANAT} (Castro-Tirado, Brandt,\& Lund 1992; Castro-Tirado \etal\
1994). Later, Mirabel \& Rodr\'{\i}guez (1994) detected from this
source the first superluminal motion in our Galaxy. By assuming a
distance of 12.5$\pm 1.5$ kpc estimated from H~I absorption, the
kinematics of the two-sided jets revealed its intrinsic velocity of
(0.92$\pm$0.08)$c$ and the inclination of 70$\pm$2$^\circ$ (Mirabel \&
Rodr\'{\i}guez 1994). The above distance has been confirmed by Chaty
\etal\ (1996) with millimeter observations and revised by Dhawan,
Gross, \& Rodr\'{\i}guez (2000a) to be 12$\pm$1 kpc from an updated H~I
spectrum, although the uncertainty may be larger than 1.5 kpc (see
Chaty \etal\ 2001). Similar jet parameters (0.98$c$ and 66$^\circ$ at
11 kpc) were obtained from other ejection events in 1997 October /
November (Fender \etal\ 1999). Recently, Greiner \etal\ (2001b) have
identified the companion to be a K-M~III star from the infrared
spectroscopic observations, classifying \grs\ as a low mass X-ray
binary, as earlier proposed by Castro-Tirado, Geballe \& Lund
(1996).The mass of the black hole has been determined to be 14$\pm$4
\solarmass\ (Greiner, Cuby, \& McCaughrean 2001a). In X-rays, \grs\
shows dramatic temporal/spectral variations occurring in regular
cycles (e.g., Belloni \etal\ 2000). Simultaneous observations in
radio, infrared, and X-ray bands indicated a close connection between
accretion disk instabilities and ejection of plasmoids emitting via
synchrotron radiation (e.g., Pooley \& Fender 1997; Mirabel \etal\
1998; Eikenberry \etal\ 1998; Klein-Wolt \etal\ 2002).

In this paper, we report results of a multiwavelength campaign of
\grs\ performed from 2000 April 16 to April 25.  This is one of the
largest coordinated set of observations ever performed for this
source, covering the wide energy band in radio ($\lambda=$13.3--0.3
cm), infrared (\J , \H , \K ), X-rays and Gamma-rays (from 1 keV to 10
MeV), during a period of 9 days in total. Results of the \chandra\
HETGS observation made on 2000 April 24 are reported by Lee \etal\
(2002), which are not included here. We report a summary of the
coordinated observations in \S~2. Then, data analysis and highlights
of the results obtained from this campaign are presented in \S~3,
followed by discussion on three topics in \S~4. In \S~5, we summarize
the conclusion.

\section{Observations}

We conducted the multiwavelength campaign when the \asca\ satellite
performed a long look at \grs\ from April 17.5 to April 25.5 (UT
throughout the paper). Simultaneous observations were made with 
{\it Rossi X-ray Timing Explorer} (\rxte )
on April 17, 20, 22, 23, 24, and 25, with \sax\ from April 21 to 24, and with
OSSE / the {\it Compton Gamma Ray Observatory} (\cgro ) from April 18
to 26. From ground facilities, radio observations were made at the
Very Large Array (VLA)
\footnote{The NRAO VLA is a facility of the National Science
Foundation, operated under Cooperative Agreement by Associated
Universities, Inc.} (observers: Mirabel and Dhawan) on April 17, 22,
and 23, and at the Nobeyama Millimeter Array (NMA) from April 17 to 19
(Kohno), and infrared observations at the Calar Alto Observatory from
April 18 to 25 (Castro-Tirado and S\'anchez-Fern\'andez) and at the
European Southern Observatory (ESO)\footnote{Based on observations
collected at the European Southern Observatory, Chile (ESO No
65.H-0247).} from April 17 to 25 (Chaty). In addition, monitoring data
of the Green Bank Interferometer (=GBI, 2.25 GHz and 8.3 GHz), All Sky
Monitor (ASM) / \rxte\ (1.5--12 keV), and BATSE / \cgro\ (20--100 keV)
were available. Rodr\'{\i}guez \etal\ (2002a) report timing analysis
from \rxte\ observations on April 17, 22, and 23, and Feroci \etal\
(2002, in preparation) report the \sax\ results. More detailed
analysis of the infrared \Kp\ data taken at Calar Alto is reported in
S\'anchez-Fern\'andez (2002). Below, we summarize observations and
data reduction for each observation. Table~1 gives a log of all the
observations.

\placetable{tbl-1}

\subsection{\cgro\ Observations}

\subsubsection{OSSE}

The OSSE instrument on the \cgro\ observed \grs\ over the period April
2000 18.7--25.6. OSSE covered the low-energy gamma-ray band, from 50
keV to 10 MeV with good spectral resolution. It had a 3.8$^\circ$
$\times$ 11.4$^\circ$ field of view (at 50 keV). Energy spectra were
accumulated in an alternating sequence of 2-minute measurements of
source and background fields (Johnson \etal\ 1993).  For these
observations, the long axis of the collimator was inclined with
respect to the Galactic plane, and the background fields were chosen
to minimize and compensate for any contamination from Galactic diffuse
emission.

The observations discussed here were performed late in the \cgro\
mission, following the last altitude reboost. At this high altitude,
the OSSE instrument was subject to additional activation in the South
Atlantic Anomaly (SAA). The radioactive decay of $^{128}$I in the
scintillator is the dominant source of internal background in the
$\sim$ 1--2 MeV band, which results in a flattening of the sensitivity
curve and --- if the background subtraction is inadequate --- can
create positive or negative spectral features in this band.

\subsubsection{BATSE}

The BATSE experiment onboard \cgro\ (Fishman \etal\ 1989) was used to
monitor the hard X-ray emission from \grs . The BATSE Large Area
Detectors (LADs) can monitor the whole sky almost continuously in the
energy range of 20 keV--2 MeV with a typical daily 3 sigma sensitivity
of better than 100 mCrab. Detector counting rates with a timing
resolution of 2.048 seconds are used for our data analysis. To produce
the \grs\ light curve, single step occultation data were taken using a
standard Earth occultation analysis technique used for monitoring hard
X-ray sources (Harmon \etal\ 1992).  Interference from known bright
sources were removed. A spectral analysis of the BATSE data indicated
that the data were well fit by a power law with a spectral index of 
--2.8.  The single occultation step data were then fit with a power law
with this index to determine daily flux measurements in the 20--100
keV band.

\subsection{\rxte\ Observations}

During the campaign, 8 pointing observations were made by \rxte\ on
2000 April 17, 20, 22, 23, 24, and 25, covering the 2--60 keV band
with the Proportional Counter Array (PCA: Jahoda \etal\ 1996) and
15--250 keV band with the High Energy X-ray Timing Experiment (HEXTE:
Rothschild \etal\ 1998).  For data reduction and analysis, we used the
FTOOLS V5.0 package provided by GSFC/NASA. The good time intervals
(GTIs) of the PCA data were selected on the two criteria: (1) the
elevation angle was 10$^{\circ}$ or larger, and (2) the offset
pointing was less than 0.02 $^{\circ}$. For spectral analysis, we used
the standard2 mode data, which has 129 energy channels with a time
resolution of 16 s. We used only events of the top layer from PCU0,
PCU2, and PCU3, which were always on during the campaign except for
April 25.24--25.26, when PCU3 was temporarily turned off. The
background spectrum was constructed from the standard model prepared
for bright sources, which includes contribution of both the cosmic
X-ray background and particle background. The HEXTE data were selected
with the same GTIs as the PCA. We used the standard mode data and
subtracted the background taken from the rocking motion. Data of
cluster-0 and cluster-1 are summed. Besides the pointing data with the
PCA and HEXTE, we also used the ASM data in the 1.5--12 keV
band\footnote{provided by the ASM team at MIT and NASA GSFC SOF/GOF
through the web}.

\subsection{\sax\ Observations}

\sax\ joined the campaign as a part of the on-going program for Target
of Opportunity observations on this source. The Narrow Field
Instruments (NFI, 0.1--300 keV, Boella \etal\ 1997a) observed the
source from April 21 19:09 (MJD 51655.798) to April 24 11:32 (MJD
51658.480) for a net exposure of 80 ks in the Medium Energy
Concentrator Spectrometer (MECS, 2--10 keV, Boella \etal\ 1997b) and
76 ks in the Phoswich Detection System (PDS, 15--300 keV, Frontera
\etal\ 1997). The two large data gaps are due to the temporary
suspension of the BeppoSAX operation during the nights of the
week-ends occurred in 2000.  Unfortunately, this limitation to the
satellite operations prevented the simultaneous observations with some
ground-based telescopes.  Standard data reduction and cleaning
procedures (Fiore \etal\ 1999) were applied in order to extract the
final scientific products. The source appears bright during the entire
observation, with an average count rate of about 100 counts s$^{-1}$
in the MECS instrument (2--10 keV) and 30 counts s$^{-1}$ in the PDS
(15--300 keV).

\subsection{\asca\ Observations}

We observed \grs\ with the \asca\ satellite (Tanaka, Inoue \& Holt
1994) in the 0.7--10 keV band from April 17.5 to 25.5. A net exposure
of 271 ksec is achieved, the longest continuous observation performed
with \asca\ for this source. \asca\ carries four X-ray telescopes,
coupled with the two Gas Imaging Spectrometers (Ohashi \etal\ 1996)
and two Solid-state Imaging Spectrometers (Yamashita \etal\
1999). Throughout the observations, the GIS was operated in the
standard pulse height mode with the nominal bit assignment, and the
SIS was operated in the 1-CCD Bright mode.

In this paper, we present only results of the GIS data, since
extremely careful treatment is required for the SIS data, which
suffered from severe pile-up as well as degradation of the response
due to integrated radiation damage at the end of the \asca\
missions. We only use data taken in the high bit-rate telemetry mode,
except when examining the presence of flare events from light curves
(see below). In the medium bit-rate, dead time becomes significant and
monitor counts measured by non-reset counter (such as L1) sometimes
spilled over due to high count rate from the source, which makes it
difficult to apply dead-time correction. The GIS data were selected
with the criteria of (1) the elevation angle of the source from the
Earth limb was higher than 5 degrees and (2) the satellite was not in
the SAA. This caused data gaps with a period of 90
minutes due to the satellite orbit. We calculate the GIS spectra using
events in a circular region within a radius of 6$'$ centered at the
peak. The energy scale of each detector were finely tuned using the
instrumental structure of the Gold M-edge and Xenon L-edge. We
estimate the resultant accuracy of the absolute gain to be about
0.5\%.

\subsection{Near-Infrared Observations at the ESO}

Near-infrared observations of \grs\ were performed with the 3.58 m New
Technology Telescope (NTT) at the European Southern Observatory (ESO)
at La Silla, Chile, during April 19--25, 2000. The telescope was
equipped with the infrared spectrograph and imaging camera Son of
ISAAC (SOFI). The observations were performed in the course of an
on-going program of Target of Opportunity observations on Galactic
hard X-ray sources (PI: Chaty).

\subsubsection{Imaging}

In imaging the broad band filters \J , \H , and \Ks\ were used, in
combination with the Large Field, giving a 4.9$\times$4.9 arcmin$^2$
field of view, with a plate scale of 0.292 arc-seconds pixel$^{-1}$.
The J-band filter is centered on 1.247 $\mu$m, with a width of 0.290
$\mu$m, representing 23\% of the wavelength, the \H -band filter is
centered on 1.653 $\mu$m, with a width of 0.297 $\mu$m, representing
18\% of the wavelength, and finally the \Ks -band filter is centered
on 2.162 $\mu$m, with a width of 0.275 $\mu$m, representing 13\% of
the wavelength.

On each night of the April 19th, 20th, 21st and 25th, the \J , \H ,
and \Ks\ band images were acquired with an integration time of 60s:
four frames of 15s each were averaged for the \J\ and \H , and six
frames of 10s each for the \Ks . The combined magnitudes given for
these nights in Table~2 are the result of 9 co-added and median
filtered 60s exposures, with random offsets and direction between
each exposure. The conditions were photometric for most of the
campaign, the seeing being typically 0.8 arcsec.

Concerning the nights of the 23rd and 24th, we continuously observed
in the \Ks\ band, taking 60 exposures with an integration time of 60s
each, by averaging 6$\times$10s frames, and randomly offseting in distance
and direction between each frame, during nearly 1.5 hours. The combined
magnitudes given for these nights in Table~2 are the result of the 60
co-added and median filtered 60s exposures. A total of 229 60s-frames
were acquired during those 6 nights.

The images were processed using IRAF reduction software. Each of the
images were corrected by a normalized dome-flat field, and
sky-subtracted by a sky image created from combining with a median
filter a total of 9 consecutive images.  The data were then analyzed
using the IRAF reduction task ``apphot'', taking different apertures
depending on the photometric conditions of the night.

Absolute photometry was performed using 2 standard stars from the new
system of faint near-infrared standard stars (Persson \etal\ 1998):
No.\ 9164 (HST P565-C) and 9178 (HST S808-C).  Each exposure of these
standard stars is the average of 7$\times$1.2s integration time frames, and
this is repeated 5 times by offsetting the images of 1 arcmin to the
North-West, North-East, South-East and South-West from the central
position, and the final image is the co-add and median filter of those
individual frames.  The Zero-points in \J ,  \H , and \Ks\ during 19--25th of
April were respectively 1.877$\pm$0.008, 2.096$\pm$0.040 and
2.652$\pm$0.062 magnitudes.

\subsubsection{Spectroscopy}

On April 22, we also performed some low resolution ($R$=600) 1.53--2.52
$\mu$m wavelength range spectroscopy with a fixed width slit of 1
arcsecond, using the red grism at 1st order, giving a dispersion of
10.22, and a resolution of 1000.

We acquired 10 exposures of 2$\times$120s each on \grs , giving a
total exposure time of 40 minutes. We also acquired 8 exposures of
3$\times$10s each on the spectrophotometric standard Hip 95550, giving a
total exposure time of 4 minutes.  We corrected the images by a
dome-flat field, and sky-subtracted the spectra by offset spectra. We
wavelength-calibrated the images thanks to a Xenon lamp spectrum
acquired with the red grism and the same setup. We thereafter
normalized the spectra.

\subsection{Near-infrared Observations at the Calar Alto Observatory}

Infrared \Kp\ observations of GRS 1915+105 were performed with the
1.23m Telescope at the German-Spanish Observatory at Calar Alto,
Spain, during April 17--24, 2000.  The telescope is equipped with the
{\sl f}/8 Mpi F{\"u}r Astronomie General-Purpose Infrared Camera
(MAGIC) which provides a 5-arcmin field of view, at a resolution of
1.2 arcsec pixel$^{-1}$.  The \Kp\ filter has a central wavelength of
2.10 $\mu$m with a width of 0.34 $\mu$m and a quantum efficiency of
about 0.6.  The seeing was typically 1.5 arcsec.  On each night,
series of nine 45s (co-added) exposures were obtained, with offsets of
$\sim$15 arcsec between each exposure.  One whole night and two
half-nights of the campaign were lost due to bad weather conditions.
A total of 924 good frames were obtained.

The images were processed using IRAF reduction software.  For each of
the images, a median-combined sky image was created from a total of 9
consecutive images, time-centered in the image to be sky-subtracted.
The resulting sky-subtracted images were then flat-fielded with an
image constructed from the difference of dome flats obtained with the
flat field lamps on and off.

The data were then analyzed using the IRAF reduction task ``apphot'',
with different apertures depending on the photometric conditions of
the night. Relative photometry was performed using 5 field secondary
standard stars, calibrated during one of our previous observational
campaigns on this field.

\subsection{NMA Observations}

The Nobeyama Millimeter Array observations of 3~mm continuum emission
toward \grs\ were made on 2000 April 17, 18, and 19. The NMA consists
of six 10 m antennas equipped with cryogenically cooled SIS tunerless
receivers. The system noise temperatures toward the zenith were about
120 K in double side band for the first two days, yet increased to
about 200 K in the last day due to a poor weather condition.  The
array was in D configuration (the most compact) for this period, and
the resultant synthesized beam were about $8" \times 6"$ at the
observing frequencies i.e., 88.6 GHz for the lower side band, and
100.6 GHz for the upper side band.  The side band separation was
achieved by 90 degrees phase switching to the reference signal.  The
backend was the Ultra Wide-Band Correlator (UWBC), which is an XF type
digital-spectrocorrelator with a bandwidth of 1024 MHz (Okumura \etal\
2000).  The passband across the channels were calibrated by
observations of 3C454.3, and the quasar B1923+210, which was about 3.5
Jy during the observations, was observed every 20 minutes to
calibrate the temporal variation of the visibility amplitude and
phase. The uncertainty in the absolute flux scale is estimated to be
about $\pm$10 \%. The raw visibility data were reduced with the
UVPROC2 developed at NRO (Tsutsumi \etal\ 1997), and then Fourier
transformed and deconvolved using a CLEAN technique implemented with
the NRAO AIPS (Astronomical Image Processing System). Upper and lower
side band images were averaged to enhance signal-to-noise ratios.
The achieved rms noise levels after the averaging were 1.3, 1.1, and
2.1 mJy for the three observing runs, respectively (indicated by the
error bars in Figure~1).

\subsection{VLA Observations}

The VLA is a multi-frequency, multi-configuration,
aperture synthesis imaging instrument, consisting of 27 antennas of
25m diameter. The receivers at 5.0, 8, and 22 GHz, were used in these
observations, with 2 adjacent bands of 50 MHz nominal
bandwidth processed in continuum mode. To get simultaneous time
coverage at 3 wavelengths, we used the antennas in 3 subarrays with
11, 7 and 9 antennas respectively.  The corresponding 1-sigma
sensitivities in 10 minutes are 0.25, 0.15, and 0.15 mJy respectively
for the 3 subarrays. (The rms noise is indicated by the error bars
on the plots of Figure~2).
The array configuration is varied every 4 months to cycle between 4
sets A, B, C, D, with maximum baselines of about 36, 11, 3.4 and 1 km.
The C array was in use during these observations.

For all observations reported here, \grs\ was unresolved by the
synthesized beams ($0.9$ arcsec at $\lambda=$1.3 cm, $1.7$ arcsec at
3.6 cm, and $4.8$ arcsec at 6.0 cm) for any subarray and
wavelength. Other sources (HII regions, see e.g., Figure~4 of 
Rodr\'{\i}guez \etal\ 1995 and Figure~1 of Chaty \etal\ 2001) are present in
the field of view, mainly at 1.4 GHz. These are well separated in the
images, and also are much weaker and resolved out by the narrower beam
at short wavelengths, so there is no confusion with \grs .

The primary flux density calibrators were 3C286 (1331+305) and 3C48
(0137+331) and the phase calibrator was 1925+211. Calibration and
imaging were carried out with standard tasks in the NRAO AIPS package.
In practice, the flux density errors are set not by the rms receiver
(thermal) noise stated above, but by errors in the flux density scale,
estimated to be 3--5\% of the measurement; and/or source variability,
depending on the occasion.

\subsection{Monitoring Observations by the Green Bank Interferometer}

The radio fluxes at 2.25 GHz ($\lambda$=3.6 cm) and 8.3 GHz
($\lambda$=13.3 cm) were monitored by the Green Bank Interferometer
(GBI) on a 2.4 km baseline with a band width of 35 MHz for each
frequency. Detailed description of the flux calibration can be found
in Foster \etal\ (1996 and references therein). Up to 12 scans were
performed every day with a 10--15 minutes integration time. The random
noise (1$\sigma$) is estimated to be about 4 mJy at 2.25 GHz and 6 mJy
at 8.3 GHz when the flux was below 100 mJy, and the data are
noise-dominated below 15--20 mJy. The data on 2000 April 21 through 23
are not available due to holiday weekend shutdown. According to the
operational notes rain may affect the 8.3 GHz data on April 17.

\newpage
\section{Data Analysis and Results}
\subsection{Multiwavelength Light Curves}

Multiwavelength light curves obtained from the whole observations are
shown in Figure~1, sorted by wavelengths. In this figure, data with
high time resolution are merged into longer time bins. For later
discussions, we plot expanded light curves with higher time resolution
in Figure~2 (PCA, GIS, and VLA on 2000 April 17.53--17.65), Figure~3
(PCA, GIS, and infrared (\Kp ) on April 19.95--20.25), Figure~4 (PDS
and MECS on April 23.78--23.83), and Figure~5 (GIS and infrared
($H$-$J$-$Ks$) on April 25.35--25.41). Below, we summarize the overall
behavior of the source during the campaign.

\subsubsection{Soft X-rays} 

It is immediately noticed from Figure~1 that before UT$\simeq$2000
April 20.3 soft X-rays ($<$25 keV) showed rapid ``flares'' or
``oscillations'' in the form of quasi-periodic ($\simeq$45 minutes)
flare-dip cycle. Detailed analysis using the 1-sec resolution data of
the \rxte\ PCA revealed that the flare phase corresponds to Class~$\alpha$ 
defined by Belloni \etal\ (2000). After April 20.3, the
source entered into a more stable state, classified as Class~$\chi$
(also referred as the low/hard state or the ``plateau'' state in
the literature). Even after April 20.3, however, a few sudden,
separated flares were still observed. Their epochs are marked by the
arrows in the GIS light curve of Figure~1. Here, to search for as many
flares as possible, we also utilized GIS LD data (one of the monitor
counts; see Ohashi \etal\ 1996) in the bit-rate medium, which are not
included in Figure~1. The presence of such flares makes the definition
of the state as a function of time untrivial.

The three flares observed after April 20.3 are essentially the same
(i.e., Class~$\alpha$) as the ones observed in the flare-dip cycles at
earlier epochs. The flare that occurred at 2000 April 23.825 was
clearly detected by the \sax : a blow up of \sax\ PDS and MECS
light curves is shown in Figure~4. As noticed, this flare was detected
only in the MECS data ($<$10 keV), with no counterpart in the high
energy detector PDS. We checked for any reasonable instrumental effect
that may have caused a spurious flare like the one we observed, and
found none. Unfortunately, the flare occurred just before an
observation gap due to Earth occultation, and therefore the study of
its time history is truncated. For the portion that \sax\ could
observe, its duration was approximately 100 s and the profile seems
similar to the start of one Class-$\alpha$ event (see Figure~2 and 3;
Belloni \etal\ 2000).

The frequency of the flares changed with time, although a correct
estimate is difficult because of the incomplete time coverage due to
the data gaps. In particular, the source showed a periodic flare-dip
cycle with a roughly constant interval of about 45 minutes between
2000 April 17.5--18.05. Then, the frequency of flares decreased and
became more irregular. There is no evidence for any flare between
April 19.54--19.96 at least in the available GIS or ASM data. After
that the source became more active again as recognized from
Figure~3. The variability pattern of the flare-dip cycle also changed
with time. On April 17 (Figure~2), the duration of a dip was roughly
comparable to that of oscillation between the two dips, whereas on
April 20 (Figure~3) the duration of the dip was about 3 times longer
than that of the flare. As we will show it later, the spectral and timing
properties in the dip phase during the flare-dip cycle observed in the
earlier epochs are very similar to those in the quasi-steady,
``plateau'' state. Hence, we may regard that during this campaign the
system was basically in the plateau state, occasionally exhibiting a
single class of flares with different rates.

\subsubsection{Hard X-rays and Gamma-rays}

Throughout the observations \grs\ was quite bright in hard X-rays: the
BATSE flux of $\sim$0.10 photons cm$^{-2}$ s$^{-1}$ (20--100 keV) was
at a typical level of the ``plateau'' state (Foster \etal\ 1996). The
OSSE flux (50--300 keV) showed a significant variability at a 20\%
level (peak-to-peak) on hours to days. After April 21 when rapid soft
X-ray oscillations ceased, the daily averaged flux increased by about
20\%. Then, it once decreased at UT$\simeq$2000 April 24.5 but
recovered at April 25.0 accompanied with decrease of the soft X-ray
flux in the 0.7--10 keV band. On much shorter time scales, hard X-ray
fluxes (above 20 keV) is much less variable than soft X-rays during
the flare-dip cycle (\S~3.5.2; see also Rodr\'{\i}guez \etal\ 2002a).

\subsubsection{Radio and Infrared}

The source showed weak flares in the radio and infrared bands
associated with the soft X-ray oscillations on April 17 and 20
(Figure~2 and 3). On April 25, a sinusoidal flux variation, which may
be regarded as a flare, was found in the combined infrared light curve
(Figure~5). We present the details in \S~3.3 and discuss the origin of
these radio/infrared flares in \S~4.1. The GBI data also show evidence
for such faint flares (e.g., on April 19). No major radio flares
exceeding $\sim$100 mJy was detected.

The \Kp\ and \Ks\ magnitudes observed during the quasi-steady state is
in the range of 13.2--13.6, which is significantly brighter than the
faintest level of $K$=14--14.5 ever observed (Chaty \etal\ 1996;
Mirabel \etal\ 1998). This result indicates that the source was quite
active throughout the campaign, even when no apparent flare was
observed. Finally, as noticed from Figure~1, the persistent infrared
flux levels seemed to correlate with the soft X-ray intensity on time
scale of hours (see \S~3.4.2 and \S~4.2).

\subsubsection{Temporal State of the Source}

To summarize, the source was predominantly in the so-called
``plateau'' or low/hard state (Class~$\chi$, or State C, in Belloni
\etal\ 2000) with a flat radio-mm-(infrared) spectrum (see Figure~6)
during the campaign. Recently, it is revealed by VLBA that this state
is accompanied with compact jets of $\sim$10 AU size (Dhawan, Mirabel
\& Rodr\'{\i}guez 2000b).
The source sometimes showed soft X-ray flares (oscillation) classified
as Class~$\alpha$. In particular, before April 20, a rapid, quasi
periodic flare-dip cycle was observed, being associated with faint
radio and infrared flares.

In the subsequent sections, we present the detailed results of (1) the
whole spectral energy distribution (SED) of \grs\ in the plateau
state, (2) the radio and infrared flares, (3) the quasi-steady
infrared emission, and (4) the high energy spectra. The last three
topics are further discussed in \S~4.

\subsection{The Spectral Energy Distribution of \grs\ in the Plateau State}

In this subsection we present the whole spectral energy distribution
of \grs\ determined from our data, to have the overall idea about
emission mechanisms at each wavelength. Here we focus on the spectra
in the plateau state (i.e., a quasi-steady state) to avoid complexity
with time variability. Note that this state is most commonly observed
from this source (Belloni \etal\ 2000) being accompanied with the
AU-scale (steady or variable) jet and is a potentially very important
phase to understand the mechanism of superluminal jets because the
state appears to be present as a precursor of large ejection events
(Foster \etal\ 1996). Figure~6 shows the energy spectrum of \grs\ in
$\nu I_{\nu}$ over the wavelength range of 12 orders of
magnitudes. They are corrected for an interstellar absorption (or
extinction): we assume $A_J$ = 7.1$\pm0.2$, $A_H$ = 4.1$\pm0.2$, and
$A_{K}$ = 3.0$\pm0.1$ from Chaty \etal\ (1996), which values are
derived from mm observations and consistent with X-ray observations of
the source. The GIS-PCA-HEXTE spectra are calculated from the data of
April 22.4--22.6 and the OSSE spectrum is from April 21.0--25.6. (The
normalization of the OSSE spectrum is multiplied by a factor of 1.35
to match the GIS one.) Detailed spectral models used to calculate
unfolded spectra are described in \S~3.5.2. For the radio and infrared
data, we plot the minimum and maximum flux densities obtained during
April 21.0--26.0, when no obvious flares were detected. As for the
flux density at 3.2 mm, we use the NMA data taken on April 16, 17, and
18, which may contain a contribution of flares.

The emission in the X-ray to Gamma-ray bands is dominated by
Comptonization, which is approximated with a power law with different
slopes below $\simeq$6 keV and above $\simeq$50 keV. In the soft band,
there is a contribution of thermal emission from the optically thick
accretion disk, which is rather small in the plateau state as widely
reported in the literature (Belloni \etal\ 2000; Muno \etal\
2001). The estimated contribution of the reflection component and of
the disk component (modeled by the Multi-Color Disk (MCD) model, Mitsuda
\etal\ 1984) are plotted separately from the best-fit parameters of
Table~4.

The emission in the infrared band has, as we will show it later, at
least 3 different components: a blackbody emission from the companion
star, a reprocessing of X-rays in outer parts of the accretion disk,
and a synchrotron emission from quasi-steady compact jets, whose
contributions to the total are estimated at 20--40\%, 20--30\%, $<$
60--30\%, respectively (see \S~4.2). In the figure we plot the
contribution from the companion star assuming a blackbody temperature
of 4800 K and a \K\ magnitude of 15.0 (Greiner \etal\ 2001b). We can
see that a contribution of the multi-color disk component itself whose
parameters are determined from the X-ray data is negligible to the
observed infrared flux, as far as the standard disk is assumed.

In the radio band, we see a flat spectrum ($\alpha \simeq 0$ for
$I_{\nu} \propto \nu^{\alpha}$). The emission is most likely
attributed to an optically-thick synchrotron radiation from compact,
quasi-steady jets. Fitting to the mean VLA fluxes at 1.3 cm, 3.6 cm,
and 6.0 cm yields $\alpha=$ $-0.03\pm 0.09$, $0.04\pm 0.03$, and
$-0.12\pm 0.07$, on April 21, 22, and 23, respectively. The best-fit
power law spectrum for the April 22 data, which are simultaneous ones
to the presented X-ray spectrum, is plotted in Figure~6. The
extrapolation of this spectrum toward shorter wavelengths roughly
agrees with the third component of the infrared emission (see
above). This is consistent that the flat synchrotron spectrum
continues, at least, till the \K\ band (1.4$\times 10^5$ GHz) (Ogley
\etal\ 2000; Fender 2001).  The absolute radio fluxes showed daily
variations, at least by a factor of 3, indicating different jet
activities on a time scale of a day but stable on a time scale of
hours.

It would be interesting to compare the SED of \grs, a {\it
microquasar}, with those of {\it quasars}. According to typical SEDs
of AGNs presented by Elvis \etal\ (1994) in their Figure~1 and
Figure~10, the ratio of $\nu I_{\nu}$ between the radio ($\nu =
10^{10}$ Hz) and X-ray ($10^{18}$ Hz) bands is roughly $\sim 10^{-2}$
for radio-loud quasars and $\sim 10^{-5}$ for radio-quiet quasars. (If
we use fluxes in UV bands instead of X-rays as a measure of accretion
power, where an optically-thick thermal emission from the accretion
disk has a peak, the ratio becomes $\sim 10^{-3}$ and $\sim 10^{-6}$,
respectively.) On the other hand, Figure~6 shows that the
radio-to-X-ray ratio is about $10^{-7}$ for \grs\ in the plateau state
(and at most $10^{-5}$ even when large radio flares reaching $\sim$1
Jy were observed). This indicates \grs\ should fall in the regime of
``radio-quiet'' microquasars. The flat radio spectrum in the plateau
state resembles that of radio-quiet quasars below $\sim 10^{11}$
Hz. However, in the case of \grs\ the flat spectrum seems to continue
till $10^{14}$ Hz, while radio-quiet quasars have the so-called
mm-break around $10^{12}$ Hz, above which the spectrum steepens
($\alpha \simgt -1$). The reason is unclear but the difference should
be taken into account when applying the same jet model developed for
AGNs to \grs , as discussed by Fender \etal\ (2000).

\subsection{Radio and Infrared Flares} 

In three epochs we detected faint flares in the radio and infrared bands
with a duration of 10--20 minutes, although none of them was
simultaneously observed in radio and infrared bands,
unfortunately. They are most likely attributed to the ejection of jets
emitting via synchrotron radiation. We present these results in
association with the simultaneous X-ray light curves. Detailed
discussions on the origin of these events are made in \S~4.1.

Figure~2 shows the VLA light curves and simultaneous X-ray light
curves on 2000 April 17 taken with the PCA and GIS. As noted, the
radio fluxes were variable on a time scale of about 45 min, with the
short wavelength varying earlier and more than the longer wavelengths. 
This behavior has often been seen previously during the X-ray
dip/flare cycles of about 30--40 min period 
(Pooley \& Fender 1997; Eikenberry \etal\ 1998; Mirabel \etal\ 1998)
and is probably due to the injection of plasma into the flat-spectrum
optically thick AU scale jet in the plateau state.
The X-ray counts show flare-dip cycles, which are probably caused by
the associated disk instability (e.g., Belloni \etal\ 1997).  Although
the time resolution is limited, we see flux enhancement in the NMA 3
mm data on April 18 and in the GBI 3.6 cm data on April 19, which
could be attributed to similar radio flares.  On the other hand, the
VLA data on April 21, 22, 23 indicate that the source is weaker and
the variability is not much more than the random noise. During these
epochs X-ray light curves did not show oscillations. The radio
spectrum of April 17 was also ``flat'' i.e., showing approximately the
same radio flux density at 6, 3.6 and 1.3 cm, at the peak of the
variation cycle. This is expected since the source size inferred from
the variability ($\simeq$1 minute) is small, and thus optically thick. 
(The optically thin, steep spectrum ejecta, if present at all, would
not be resolved by the large beam $\geq 1"$).

Similar faint flares were also detected in the \Kp\ band on April
20. Figure~3 shows a blow up of the \Kp\ band light curve at 1 minute
resolution, plotted with the \asca\ GIS and \rxte\ PCA counts. The
X-ray flares correspond to the same class (Class~$\alpha$) as April
17, although the whole variability pattern seems slightly different
from the previous epoch in the sense that the duration of the dip
relative to the flare is longer and that the amplitude of the first
peak is more prominent compared with the subsequent peaks. As seen
from the figure, four (or five if the beginning of the data is
counted) separated infrared flares are detected, each has a
peak-to-peak amplitude of about 0.3 magnitudes (about 20 mJy if
dereddened). Because of the gaps in the GIS data, we could detect only
one peak that marks the beginning of the X-ray flare at April
20.07. Nevertheless, the linear rise patterns of the X-ray intensity
suggest the presence of such flares between them, supporting
one-to-one correspondence between the infrared and X-ray flares. For
later discussion, we estimate the timing of the flares that were not
directly observed. In the last panel of Figure~3, the estimated
positions of the first peak in each oscillation phase are marked by
the downward arrows with a larger width representing the probable
error (those with a smaller one are actually observed peaks). Here we
have assumed, based on the variability pattern observed in the GIS and
PCA data around this epoch, that (1) the duration of the oscillation
phase is 15 minutes, and that (2) the flux level at the bottom of the
dip, just following the oscillation phase, can be known by a linear
interpolation from those before and/or after the cycle.

Figure~5 shows combined infrared light curves taken at the ESO on April 25,
together with the GIS light curve. The first 8 points correspond to
the \J\ band data, next 9 ones to \H , and the rest to \Ks . To make
it easy to see the variability over the whole period, we add offsets
to the \J\ and \H\ magnitudes to plot them at the same level as \Ks ,
so that the data points are smoothly connected by a linear
extrapolation. The light curve suggests the presence of a significant,
sinusoidal variability with a period of about 10 minutes with a
peak-to-peak amplitude of about 0.5 magnitude. Although,
unfortunately, most of the strictly simultaneous epoch is not covered
by the GIS, there is no evidence for any X-ray flare around this time
within offsets of 20--30 minutes.

\subsection{Quasi-Steady Infrared Emission}

\subsubsection{Infrared Spectra}

The infrared spectra (1.53--2.52 $\mu$m) taken on 2000 April 22 with a
resolution of 600 are shown in Figure~7. The visible lines, as already
seen by e.g., Mart\'{\i} \etal\ (2000) and Greiner \etal\ (2001b) are
summarized in Table~3. The presence of He II is consistent with the
fact that the source was in a low state on April 22 as seen on the
X-ray and radio light curves. We also detect the $^{12}$CO and
$^{13}$CO absorption band heads, as discovered by Greiner et
al. (2001b). The presence of both He~II and $^{12}$CO/$^{13}$CO
absorption band heads being a signature of a low-mass system, this
confirms the claims of these authors that the companion star is a K-M
III star. The \K\ magnitude of the star is estimated to be 14.5--15.0
(Greiner \etal\ 2001b), which accounts for about 20--40 \% of the total
magnitude observed in the plateau state.

\placetable{tbl-2}

\subsubsection{Correlation between Infrared and X-rays Fluxes in the Plateau State}

In carefully comparing the GIS and infrared \Kp\ light curves in
Figure~1, one may notice a good correlation between the quasi-steady
flux levels in the two bands, except for the data of April 20, where
flares were observed. Such correlation is clearly seen also from the
data on April 25 alone, where both soft X-ray and infrared fluxes
gradually decreased with similar profiles. Figure~8 shows the
correlation between the \Kp\ magnitude, taken at the Calar Alto
observatory, and the corresponding GIS count rates (0.7--10 keV) using
exactly simultaneous exposures. Different symbols correspond to
different days. Sixteen data points are merged into one point for the
\Kp\ magnitude, thus giving averaged fluxes with a time resolution of
about 16 minutes. We do not include the \Ks\ data taken at the ESO,
for which available simultaneous GIS data are limited, to avoid
systematic errors by instrumental effects.

Figure~8 demonstrates the good correlation between them except for the
data on April 24. Such correlation can be most easily explained in
terms of reprocessing of irradiating X-rays. The reprocessor cannot be
located farther away from the black hole than $\sim 10^{13}$ cm
because the time delay is less than $\sim$10 minutes as seen from the
April 25 data, and hence it is most likely the outer parts of the
accretion disk. We discuss this correlation quantitatively in
\S~4.2. The reason for the deviation of the April 24 result is not
clear at present.

\subsection{High Energy Spectra}

\subsubsection{OSSE Spectra}

We analyze the OSSE spectrum, covering the highest energy band from 50
keV to 10 MeV. As mentioned above, \grs\ was in active states showing
rapid soft X-ray oscillations before 2000 April 21, and then became
more stable. To see if there is any difference of the high energy
spectra between before and after UT = April 21.0, we separately
calculated OSSE spectra in the two epochs, from April 18.7 to April
21.0, and from April 21.0 to April 25.6. (It is not feasible to
extract the OSSE spectra with the exactly same GTIs as the X-ray data
due to limited photon statistics.)  Assuming a power law, we find no
significant difference in terms of the photon index except that the
normalization in the second epoch is larger by about 20\%. Hence, we
sum the whole data from a total exposure of 169 ks. We find that the
whole OSSE spectrum from 0.05 to $\sim$1 MeV can be well described by
a single power law ($\chi^2/$dof=5.8/16). We obtained a photon index
of 3.12$\pm$0.06 with a flux of (2.2$\pm$0.1)$\times 10^{-9}$ \ergs\
(0.05--10 MeV; this value is not corrected for any calibration
difference from the other instruments), confirming the results of
Zdziarski \etal\ (2001).  The flux is the largest ever observed with
OSSE and the photon index is the typical value of this source (Grove
\etal\ 1998; Zdziarski \etal\ 2001). Figure~9 show the unfolded OSSE
spectrum. It is apparent that the data show no evidence of a spectral
cutoff.  The data appear to suggest a slight hardening of the spectrum
above 500 keV, but we note that the total significance of the excess
over the power law model is less than 1.6$\sigma$.  Given the
additional systematic uncertainty (not shown in the error bars) in
energies above an MeV (see the discussion in \S~2.1.1), we do not
consider the excess to be significant, and we make no claim to the
existence of any spectral hardening above the best-fit power law. If
we combine it with the \rxte\ HEXTE spectrum taken on April 22 to
cover a wider energy range over 25 keV--10 MeV, we find a spectral
break at $56\pm 8$ keV, below which the photon index is about
2.67$\pm$0.05. Note that the presence of the spectral break is
confirmed by the HEXTE spectrum alone (Figure~10), which gives a break
energy at $42^{+12}_{-5}$ keV, consistent with the above result.

\subsubsection{GIS-PCA-HEXTE Spectra}

In this section, we perform joint spectral analysis from the \asca
/GIS, \rxte /PCA, and \rxte /HEXTE data, covering the 1--200 keV band.
The detailed analysis of the spectral evolution during the flares is
beyond the scope of this paper, and we present the overall spectral
properties in typical states. In the joint spectral fitting, we paid
special attention to avoid systematic errors. We made
cross-calibration between the instruments using the spectra of Crab
Nebula and effectively corrected the quantum efficiencies in the
response matrices for the slight difference in the obtained best-fit
Crab photon index between different instruments, which is turned out
to be about 0.05 between GIS and PCA with the use of the latest
responses. Absolute normalizations presented in this paper are
determined from the GIS (which gives the 2--10 keV Crab flux of
$2.2\times10^{-8}$ \ergs ). For all the spectra, we added a 1.5\%
systematic error in each energy bin, and conservatively, 3\% for the
GIS data below 2.2 keV, considering possible uncertainties in the
responses for highly absorbed sources. We limited the energy range of
the GIS, PCA and HEXTE spectra to 1.2--10 keV, 3.5--25 keV and 20--200
keV, respectively.

As mentioned above, during the \rxte\ observations on April 17 and 20,
the source showed a peculiar pattern of variability, characterized by
a rapid oscillation lasting for typically 500--1000 sec, followed by a
decline of $\sim$ 200 sec and a slow rise of $\sim$ 1000--2000
sec. The soft X-ray intensity and the hardness have correlations: when
the intensity is high, the spectrum becomes hard in the 0.7--10 keV
range but soft in the 3--31 keV range.
We here define four characteristic states: (State~I) = the
``oscillation high'' state, when the soft X-ray flux is larger than
roughly the mean value during the oscillation phase ($>$1800 counts
s$^{-1}$ per PCU in the 16 sec resolution light curve), (State~II) =
the ``oscillation low'' state ($<$1800 counts s$^{-1}$ per PCU),
(State~III) = the ``dip'' state, when the intensity increases
linearly with time between oscillations, and (State~IV) = the plateau
state, observed after 2000 April 22.

We create the spectra separately for the 4 states defined above,
although the \asca\ and \rxte\ data are not strictly simultaneous: for
State~I, II and III, we use the \rxte\ data from the April 17
observation and the GIS data taken between April 17.51 and 18.90. For
State IV, we calculate them separately for April 22 and 23 from each
epoch covered by \rxte\ but mainly present the results from the April
22 data unless otherwise mentioned. For modeling of the continuum, we
use the multi-color disk model plus a hard component whose form
is described below, modified with the interstellar absorption. Based
on the analysis of frequency-resolved energy spectrum at frequencies
higher than several Hz done by Yamaoka (2001), we adopt a
phenomenological model for the hard component, a broken power law with a
break energy at 6--7 keV below which the photon index is fixed at
1.65, multiplied by the high energy cutoff (``highecut'' in the XSPEC
package). The cutoff is necessary to account for the spectral
break (or curvature) around 50 keV as mentioned above. Note that this
approximation is only valid for spectral fit in the range below
$\sim$200 keV because no cutoff is present at higher energies as we
have shown above.

Figure~10 shows unfolded spectra for each state. We find that the fit
with this continuum model alone is acceptable for State I and II but
not for State III and IV ($\chi^2/$dof=174/154, 178/154, 532/154, and
794/154, respectively). Nevertheless, the residuals indicates the
continuum is reproduced within a 5\% level with this model for any
state, which is sufficient to see the difference of the continuum from
these unfolded spectra. As noticed from the figure, the spectra in the
dip (State III) and plateau (IV) states are very similar to each other
and the contribution of the soft, disk component seems much weaker
than in the oscillation states (State I and II). In contrast, the
difference of the hard band spectrum above $\sim$30 keV is not
conspicuous between these states, except for a slight difference in
the normalization. We can see that a canonical ``MCD plus power law''
model cannot be used to reproduce the continuum appropriately at
least for State III and IV, because the soft X-ray emission below
$\sim$6 keV is weaker than the extrapolation of a single power law
determined in higher energy bands.

In the spectra of State III and IV, we notice a broad absorption-edge
like structure above $\sim$7 keV and a hump around 25 keV, which are
most likely due to the presence of a reflection component. This is
already discussed in Zdziarski \etal\ (2001). Alternatively, such
feature might be explained by a partial covering. In our case,
however, the derived column densities ($N_{\rm H} \simeq
3\times10^{24}$ cm$^{-2}$) exceed one Thomson optical depth, which
would lead to a significant increase of the intrinsic luminosity
corrected for Thomson scattering, for example $>2\times10^{39}$ erg s$^{-1}$ in
State~III. This is above the Eddington luminosity of the 14
\solarmass\ black hole, and hence we think the ``pure'' partial covering
case is unlikely.
We finally find that the combined GIS-PCA-HEXTE spectra in these
states can be fit with a sum of the MCD model, a broken power law
modified with the high energy cut, and a reflection component. In
addition, an absorption edge is required at 9.3 keV; this is expected
when H-like iron ions are present in the line of sight, as indicated
by the iron-K absorption line at 7.0 keV in the \chandra\ HETGS data
taken at a quasi simultaneous epoch (Lee \etal\ 2002). For the
calculation of the reflection component, we use the ``pexriv'' code
by Magdziarz \& Zdziarski (1995), approximating the incident spectrum
by a cutoff power law model with the parameters determined from the
observed spectrum for each state. Since the reflection component
should be accompanied by an iron-K emission line, we include it in the
fit, assuming a narrow one fixing its energy at 6.5 keV. The H-like
iron-K absorption line is also included with the same parameters as
Lee \etal\ (2002), which has an equivalent width of about 20 eV.

Table~4 summarizes the best-fit parameters in each state (we do not
include the reflection component or edge structure at 9.3 keV in State
I or II since they are not required). The best-fit model for the
spectra on April 20 (=State~IV) are plotted in Figure~6. We obtain the
innermost radius of the MCD component to be 30$\pm$2 km, 32$\pm$2 km,
52$\pm$3 km and 61$\pm$7 km on April 22 (49$\pm$8 km on April 23) for
State I, II, III, and IV, respectively, assuming the distance of 11
kpc and inclination of 66$^\circ$. Note that these values may be
referred only for relative estimates for the disk radius between
different states, since the absolute radius obtained from the fit
strongly depends on the modeling of the hard component in the
soft-energy range, which is highly uncertain. 

\section{Discussion}
\subsection{Origin of the Radio and Infrared Flares}

\subsubsection{A Brief Summary of Radio/Infrared Flares Previously Reported from \grs }

From \grs\ at least 3 classes of jets have been observed so far
according to the classification by Eikenberry \etal\ (2000): (Class-A)
major ejection events with radio fluxes of $\sim$ 1 Jy (Mirabel \&
Rodr\'{\i}guez 1994; Fender \etal\ 1999), (Class-B) repeated flares
with a period of 30--60 minutes with (dereddened) amplitudes of
$\sim$100--200 mJy, associated with the X-ray hard-dip/soft-flare
cycle (e.g., Mirabel \etal\ 1998; Eikenberry \etal\ 1998), and
(Class-C) faint infrared flares with amplitudes of $\sim$10 mJy
observed in the soft-dip/soft-flare cycle (Eikenberry \etal\ 2000; see
also Feroci \etal\ 1999). Interestingly, the infrared peak of these
Class-C events appeared to precede the X-ray dips by 200--600 sec
(unless there were time delays of longer than 2000 sec that varied
flare by flare). Accordingly they suggest ``outside-in'' origins for
the disk-jet connection. Note that the above classification is only
phenomenological at present and distinction between Class-B and Class-C
jets is sometimes ambiguous particularly when X-ray data are not
available.

For Class-B jets observed in the X-ray state of Class $\beta$ (Belloni
\etal\ 2000), many authors consider that they are triggered by a
``spike'' in the soft X-ray light curve separating the hard-dip and
soft-dip phases (e.g., Mirabel \etal\ 1998; Yadav 2001). The similar
decay times between the radio and infrared bands indicates that the
adiabatic cooling dominates the radiative cooling (Fender \etal\ 1997;
Fender \& Pooley 1998). As a result, intrinsic power of jets is
estimated to be much larger than that observed as radiation (by $\geq
0.05^{-1}$, Fender 2001). The frequency-dependent peak delay of the
flare peaks (higher frequencies peak earlier) can be explained in
terms of expanding plasmoids radiating through synchrotron emission
(e.g., Mirabel \etal\ 1998). However, as argued by several authors
(e.g., Fender \& Pooley 1998), the flat spectra of their peak fluxes
in the radio to \K -band is difficult to explain by a simple ``van der
Laan'' model (van der Laan 1966; Mirabel \etal\ 1998), rather favoring
the partially self-absorbed conical jet models applied for AGNs
(Blandford and K{\" o}nigl 1979). More updated models for the
quasi-steady jets are proposed by e.g., Falcke \& Biermann (1999) and
Kaiser, Sunyaev, \& Spruit (2000).

\subsubsection{Interpretation of Our Results}

The flares observed on April 17 by VLA, in particular at 1.3 cm, and
those in the \Kp\ band on April 20 are very similar in their
amplitudes, time profiles, and associated X-ray state (Class~$\alpha$ ). 
This tempts us to consider that these flares are essentially the same
class of events. The (dereddened) amplitudes, about 10--20 mJy, may
suggest their classification as Class-C jets in terms of the flux. The
fast rise time, typically $\sim 1$ min, constrains the size of the
emitting region to be less than $\sim 10^{12}$ cm. The decay time
scale of the flares (about 10 minutes) is also similar between
$\lambda=$1.3 cm and 2.2 $\mu m$, supporting previous claims that the
adiabatic cooling is dominant in the frequency range at least below
$1.4\times10^{5}$ GHz. By assuming a flat spectrum below
$1.4\times10^{5}$ GHz with an intensity of 15 mJy and the distance of
11 kpc, the power of the jet calculated according to Fender (2001)
\footnote{In the formula for $L_{\rm jet}$ given in Section 7 of
Fender (2001) $10^{36}$ should be replaced by $1.4\times10^{35}$.} 
becomes $L_{\rm jet} \geq 6\times 10^{36} (\frac{0.05}{\eta})
F(\Gamma,i) $ erg s$^{-1}$. Here $\eta$ is the ratio of the observed
power to the total power, which is estimated at $\leq 0.05$ (Fender \&
Pooley 2000), and $F(\Gamma,i)$ is the correction factor for
relativistic bulk motion.  If we assume the inclination of
$i=66^\circ$ and the velocity of 0.98$c$ (thus giving $F(\Gamma,i)
\simeq 140$) measured for the Class-A event of 1997 October/November
(Fender \etal\ 1999), the minimum jet power reaches $8\times 10^{38}$
erg s$^{-1}$. This value amounts to about twice the observed X-ray to
Gamma-ray luminosity, $4\times10^{38}$ erg s$^{-1}$ (1 keV--10 MeV) on
April 22.

Timing of the radio/infrared flares with respect to the X-ray
variability is important to constrain the mechanism of jets. For the
radio flares on April 17, the observed frequency-dependent peak delay
can be explained by an expansion of a plasmoid or an internal-shock
region (see Kaiser, Sunyaev, \& Spruit 2000), which predicts a time
delay of $\Delta t \propto \lambda^{2/3}$ (van der Laan model for an
electron energy distribution of $p\simeq0$ for $N(E) \propto E^{-p}$)
or $\Delta t \propto \lambda^{1}$ (self-similar regime of the
Blandford \& K{\" o}nigl jet), respectively. As seen from Figure~2,
$\Delta t \sim$ 5 minutes between $\lambda=$ 1.3 and 3.6 cm. In either
model, $\Delta t$ at 1.3 cm is estimated at $\sim$5 minutes,
indicating the start time of each radio flare coincides with the time
of the first peak in the X-ray oscillation phase.
The timing of radio flares on 1996 October 25 associated with 
Class~$\alpha$ (Figure~5 of Pooley \& Fender 1997) seems consistent
with this picture, although their amplitudes, about 50 mJy, were about
3 times larger than ours.
These first X-ray peaks may correspond to the soft X-ray ``spike''
of Class~$\beta$, which separates the states with hard
and soft spectra, possibly triggering Class-B jets (e.g., Yadav 2001). 
Obviously, the duration of the 1.3 cm flare is shorter
than that of the X-ray oscillation phase. This rules out the scenario
that the radio flare is a superposition of many short flares of a
similar amplitude produced by ``every'' X-ray peak during the
whole oscillation phase.

On the other hand, the interpretation for the infrared flares observed
on April 20 is not straightforward. As noticed from Figure~3, the
timing of the rise of each infrared flare does not match any of the
first peak in X-ray oscillation, which are indicated by the arrows in
the third panel. (This argument is firm, at least for the infrared
flare on April 20.08--20.09, which is free from the data gap problem.)
Therefore, if we relate the timing of infrared flares to the start
time of each X-ray oscillations, as done for the April 17 events, two
possibilities arise i.e., (1) the infrared flare {\it followed} the
first peak of X-ray oscillation phase, or (2) the infrared {\it
preceded} the X-ray (as the case suggested by Eikenberry \etal\ 2000). 
In either case, as seen from Figure~3, the time delay seems to be
different flare by flare, ranging from $\simeq$5 to 30 minutes (plus
integer times the one cycle period, $\simeq$45 minutes). We favor the
first possibility (infrared delay) because it is more natural to
attribute the ejection to change of the state of the accretion disk
observed in X-rays and because it seems difficult to find physical
mechanisms causing such a long delay in the second case. In the first
case, the longest delay time indicates the distance of the infrared
emitting region can be far by $5\times10^{13}$ cm from the black hole
assuming the jet velocity of $\sim0.9c$. This is consistent with the
order of the size of the quasi-steady jet in the plateau state,
$\sim$10 AU (Dhawan, Mirabel, \& Rodr\'{\i}guez 2000b). The highly
delayed infrared flare can then be understood by an internal shock
that occurred far away from the jet base, being triggered by a matter
ejected at the beginning of the X-ray oscillation. On the other hand,
the size of the ``lit-up'' region by the internal shock must be $\leq
10^{12}$ cm from the constraints of the time variability, which is
much smaller than the scale of jet length. This implies that the jet
is well collimated with an opening angle of $\simlt 1/10$, which is
consistent with the values derived from other observations (e.g.,
Fender \etal\ 1999; Chaty \etal\ 2001).

We consider it unlikely, though cannot rule out, that the ejections
occurred at the {\it end} of the oscillation i.e., {\it start} of the
dip. If the first two flares at April 20.08--20.09 and 20.125--20.135
had been triggered by the starts of the dip within $\simeq$20 minutes,
the infrared flare at April 20.15 would have occurred {\it before} the
start of the dip by $\simeq$5 minutes, contrary to the previous
causality. Therefore, in either case where the infrared flares {\it
followed} or {\it preceded} the start of the dip, we have to introduce
a long time delay, larger than 45 minutes, to reconcile the timing
relation for the event at April 20.15. Finally, there could be even an
extreme possibility that the ejections occurred during the dip phase
without showing no apparent signature in X-rays.  Indeed, the infrared
flare-like flux variation detected on April 25 appeared not to be
accompanied by any strong X-ray variability, although this is not
conclusive due to the data gap. The detection of such variability
might be consistent with the picture that the plateau is a quite
active state exhibiting continuous internal shocks where a weak
perturbation in the jet flow could cause faint radio/infrared flare
even without a dramatic change in the accretion state observable in
X-rays.

In summary, it is not clear whether or not all the faint ($\simlt 40$
mJy) radio/infrared flares observed so far indeed belongs to a single
class of events. Although one-to-one correspondence between
radio/infrared and X-ray variability pattern strongly indicates a
tight disk-jet connection, it is not obvious yet what kind of
transition seen in X-rays actually triggers the ejections. For
example, the events detected by Eikenberry \etal\ (2000) are
associated with the soft-dip/soft-flare cycle, while ours seems to be
associated with the transition from the hard-dip to the soft-flare
phase, similarly to the case of Class-B jets (Mirabel \etal\
1998). Even within similar states, as in our observations on April 17
and 20, the timing relation between the radio/infrared flare and
X-rays seems to be different.  More observations are necessary to
reveal the nature of these ``faint'' flares. They are potentially very
important to understand the physics of the quasi-steady jet in the
plateau state, considered to exist in many X-ray binary systems
(Fender \etal\ 2000).

\subsection{Origin of the Quasi-Steady Infrared Emission}

Here we summarize possible origins of infrared emission of \grs, as
listed in e.g., Mirabel \etal\ (1997): (1) emission from the companion
star, (2) reprocessing of X-rays in the outer parts of the accretion
disk, (3) optically thin free-free emission from an X-ray driven wind
(van Paradijs \etal\ 1994), (4) synchrotron emission from jets (Sams,
Eckard \& Sunyaev 1996), (5) Doppler-shifted line emission from ions
in the relativistic jets (Mirabel \etal\ 1997), and (6) thermal dust
reverberation of energetic outbursts (Mirabel \etal\ 1996). Below we
constrain the contribution of the second origin.

As we have shown in Figure~8, we find a good correlation between 
the flux of the quasi-steady infrared emission and that of X-rays.
To be quantitative, we fit the relation between the infrared flux
$F_{\rm IR}$ (unit: mJy) dereddened with $A=3.0$ versus the X-ray flux
$F_{\rm X}$ (unit: GIS counts s$^{-1}$) with a linear function in the form of
$$
F_{\rm IR} = F_{\rm IR0} + C \times F_{\rm X}, 
$$
based on a simple assumption that the reprocessed infrared flux was
proportional to the irradiating X-ray flux and other components were
constant. The data of April 24 are not included. The fit is quite good
and we obtain the best fit parameters of $F_{\rm IR0}$ = 37 mJy and
$C=7\times10^{-5}$. This indicates that the contribution of the
reprocessing of X-rays is about 20--30 \% of the total observed
magnitude (\Kp\ = 13.25--13.4 on 2000 April 21--25).

Thus, the contribution from the companion star and from reprocessing of
X-rays in outer parts of the accretion disk is likely to account for
about 20--40\% and 20--30\% of the observed, quasi-steady K magnitudes
in the plateau state, respectively. The rest 60--30\% of the total
infrared flux (i.e., 30--15 mJy) must have other origins than (1), (2)
and (6) listed above. As mentioned earlier, if we attribute this to
synchrotron radiation from the jets, the result is consistent with the
picture that the emission from quasi-steady compact jets continues
from the radio to near-infrared bands with a nearly flat spectrum.

\subsection{Origin of the High Energy Spectra}

We have presented the energy spectra of GRS1915+105 in the 1 keV -- 10
MeV band combining the contemporaneous \asca\ GIS, \rxte\ PCA and
HEXTE, and \cgro\ OSSE observations. These data provide the most
complete coverage in high energy bands ever presented in the
literature for \grs\ and are used to construct the spectral model
(Figure~6). In the 1--200 keV band, using the simultaneous data, we
could obtain the energy spectra separately in the three typical states
during the soft X-ray flare (Class~$\alpha$) and in the plateau state
(Class~$\chi$).

Power law spectra with photon indices of $\sim 2.5$ extending above
$\sim$1 MeV without cutoff are often observed from black hole
candidates (BHCs) in the ``soft state'' (e.g., Grove \etal\ 1998;
Gierli{\'n}ski \etal\ 1999), in which most of the energies are emitted
below $\sim$ 5 keV via thermal emission from the optically thick
accretion disk. These soft-state energy spectra are obviously
different from the typical energy spectra in the hard state, which
show power-law spectra with photon indices $1.7$ and a thermal cutoff
around 100--200 keV (e.g., Gierli{\' n}ski \etal\ 1997; Grove \etal\
1998). The fact that \grs\ has a very ``soft'' ($\Gamma\approx3$)
spectrum indicate that its spectral state corresponds to the soft
state rather than the hard state, although, in the plateau state, a
contribution of the disk component seems much weaker than in the
soft-state spectra of other BHCs. The absence of high energy cutoff
suggests the presence of non-thermal electrons, as Zdziarski \etal\
(2001) indicate that the spectra of \grs\ can be explained by
Comptonization model with a hybrid plasma with thermal and non-thermal
electrons.

The spectral change in the oscillation phase have been reported by
many authors, which is considered to be caused by thermal-viscous
instability (e.g., Belloni \etal\ 1997). In fact, as pointed by
Yamaoka (2001), the apparently high temperature and small radius in
the oscillation high state, as confirmed from our observations, can be
explained by emission from optically-thick advection dominated
accretion flow, so-called a slim-disk (Watarai \etal\ 2000), as a
consequence from such transition.  Such transition may trigger
ejection of jets observed in radio and infrared bands (\S~4.1).

We found that the spectrum in the dip phase between rapid oscillations
and that in the more stable plateau state look similar, confirming
that they are essentially the same class of spectral state (State C,
Belloni \etal\ 2000). Indeed, they also share similar timing
properties, showing QPOs with harmonics (Rodr\'{\i}guez \etal\ 2002a). 
The correlation between the QPO frequency and the disk radius in the
plateau state between April 22 and 23 is consistent with the claims by
previous authors, in the sense that a higher frequency QPO is observed
with a smaller disk radius. Such correlation could be explained by the
accretion ejection instability (Tagger \& Pellat 1999; Rodr\'{\i}guez
\etal\ 2002b; Varniere, Rodr\'{\i}guez, \& Tagger 2002).

For State III and IV, we find the solid angle of the reflector
subtending the X-ray source is $\Omega / 2 \pi \simeq$ 0.6--0.9. This
value predicts a total equivalent width of iron-K emission lines of
50--100 eV (e.g., Basko 1978; George \& Fabian 1991), whereas we
obtain only $\sim$10 eV from our fits. Also, the \chandra\ HETGS data
also show very small equivalent widths for (narrow) emission lines in
the range of 6.4--6.6 keV, which are consistent with our results. Such
apparently small line intensities could be explained by significant
blurring of the line profile via Doppler redshift and/or Compton
scattering, or other physical mechanism such as Auger destruction
(Ross, Fabian \& Brandt 1996). Note that the obtained edge depth from
H-like iron, $\tau \simeq 0.05$, corresponds to the iron column
density of log $N_{\rm Fe26} \simeq 18.7$, which is by about one order
of magnitudes larger than that derived from a simple curve of growth
analysis of the absorption line (Lee \etal\ 2002).  This implies that
the resonance absorption lines are partially refilled by the emission
line photons at the same energy.

In our observations, we find that the spectrum above $\sim$30 keV does
not significantly differ between different states. The spectrum keeps
the same slope ($\Gamma\approx3$ above 50 keV), and shows similar
fluxes even during the oscillation phase in spite of the large flux
variation below $ \simlt 10$ keV. The eight OSSE observations
summarized by Zdziarski \etal\ (2001) also show that, on much longer
time scales, the power law slope above 50 keV is almost always
constant around $\Gamma \simeq 3.0$ and the flux stays within factor
of 2 except for one occasion. Thus, the constant slope and the lack of
correlation with soft X-ray states may be persistent features of the
hard X-ray spectrum of \grs . The stable emission above $\sim 50$ keV
implies that the energy distribution of high-energy electrons
responsible for Comptonization is regulated by some (unknown)
mechanism, not being affected by the change of the state in the
accretion disk. These results give strong constraints for theoretical
modeling of the accretion flow in \grs .

\section{Conclusion}

We have performed a multiwavelength campaign of \grs\ from 2000 April
16 to 25, covering the wide energy band in radio ($\lambda=$13.3--0.3
cm), near-infrared (\J , \H , \K ), X-rays and Gamma-rays (from 1 keV
to 10 MeV), with \asca , \rxte , \sax , \cgro , and ground facilities
including the VLA, the NMA, the ESO, and the Calar Alto Observatory.
This is one of the largest coordinated observations ever performed for
this source. Main results obtained from this campaign are summarized
as follows.

(1) The source was predominantly in the ``plateau'' state (Class~$\chi$) 
and sometimes showed soft X-ray flares (oscillation) classified
as Class~$\alpha$. Before April 20.3, a rapid, quasi periodic
flare-dip cycle was observed, being associated with faint radio and infrared
flares.

(2) The spectral energy distribution of \grs\ in the plateau state is
presented covering the energy range of 12 orders of magnitudes. The
radio to near-infrared spectra are flat, consistent with the presence
of optically-thick compact jet. The X-ray to Gamma-ray emission is
dominated by Comptonization.

(3) In three epochs we detected faint flares in the radio or infrared
bands with amplitudes of 10--20 mJy. The radio flares observed on
April 17 shows frequency-dependent peak delay, consistent with an
expansion of synchrotron-emitting region starting at the transition
from the hard-dip to the soft-flare states in X-rays. On the other
hand, the infrared flares on April 20 appear to follow (or precede)
the beginning of X-ray oscillations with an inconstant time delay of
$\simeq$ 5--30 min. This implies that the infrared emitting region,
probably the place of an internal shock in the quasi-steady jet, is
located far from the black hole by $\simgt 10^{13}$ cm, while its size
is $\simlt 10^{12}$ cm constrained from the time variability.

(4) A good correlation is found between the quasi-steady flux level in
the near-infrared band and that in the X-ray band after 2000 April 21.
This indicates that a part of the observed infrared emission is
attributable to reprocessing of X-rays, probably in outer parts of the
accretion disk. The contribution from the companion star and from
reprocessing of X-rays accounts for about 20--40\% and 20--30\%,
respectively, of the quasi-steady \K\ magnitude (\Kp =13.25--13.4) in
the plateau state.

(5) The time averaged OSSE spectrum from 50 keV to $\sim$1 MeV is
represented by a single power law with a photon index of 3.1 with no
significant high energy cutoff, confirming the results of Zdziarski
\etal\ (2001). It did not show significant difference, except for the
normalization, between before and after April 21.0 when the rapid soft
X-ray flares ceased. The continuum in the 1--200 keV band can be
modeled by a broken power law with a break energy at 6--7 keV,
modified with a high energy cutoff, plus the multi-color disk
model. In addition, we significantly detect a reflection component
from a cold (or slightly warm) matter with a solid angle of $\Omega/2
\pi = 0.6-1.0$ in the spectra of the dip and plateau states.  The
power-law slope above $\sim$30 keV is found be very similar between
different states in spite of large flux variations in soft X-rays,
implying that the energy distribution of high energy electrons
responsible for Comptonization is regulated by some (unknown)
mechanism, not being affected by the change of the state in the
accretion disk.

\acknowledgments

We thank Neil Gehrels for making the OSSE observation possible
in coordination with the campaign and Robert Fender for
discussion. S.C. is very grateful to the ESO/NTT team, and especially
Leonardo Vanzi, for all the high quality service observations
performed every day during one week of this multiwavelength campaign,
as a target of opportunity. S.C. also acknowledges support from grant
F/00-180/A from the Leverhulme Trust. We also thank Kazuhiro Sekiguchi
(Subaru telescope) and Taichi Kato (variable star network) for
their efforts and help in organizing the campaign. This research has
made use of the data of the Green Bank Interferometer (GBI), which is
a facility of the National Science Foundation operated by the NRAO in
support of NASA High Energy Astrophysics programs.

\ifnum1=1

\clearpage

\figcaption[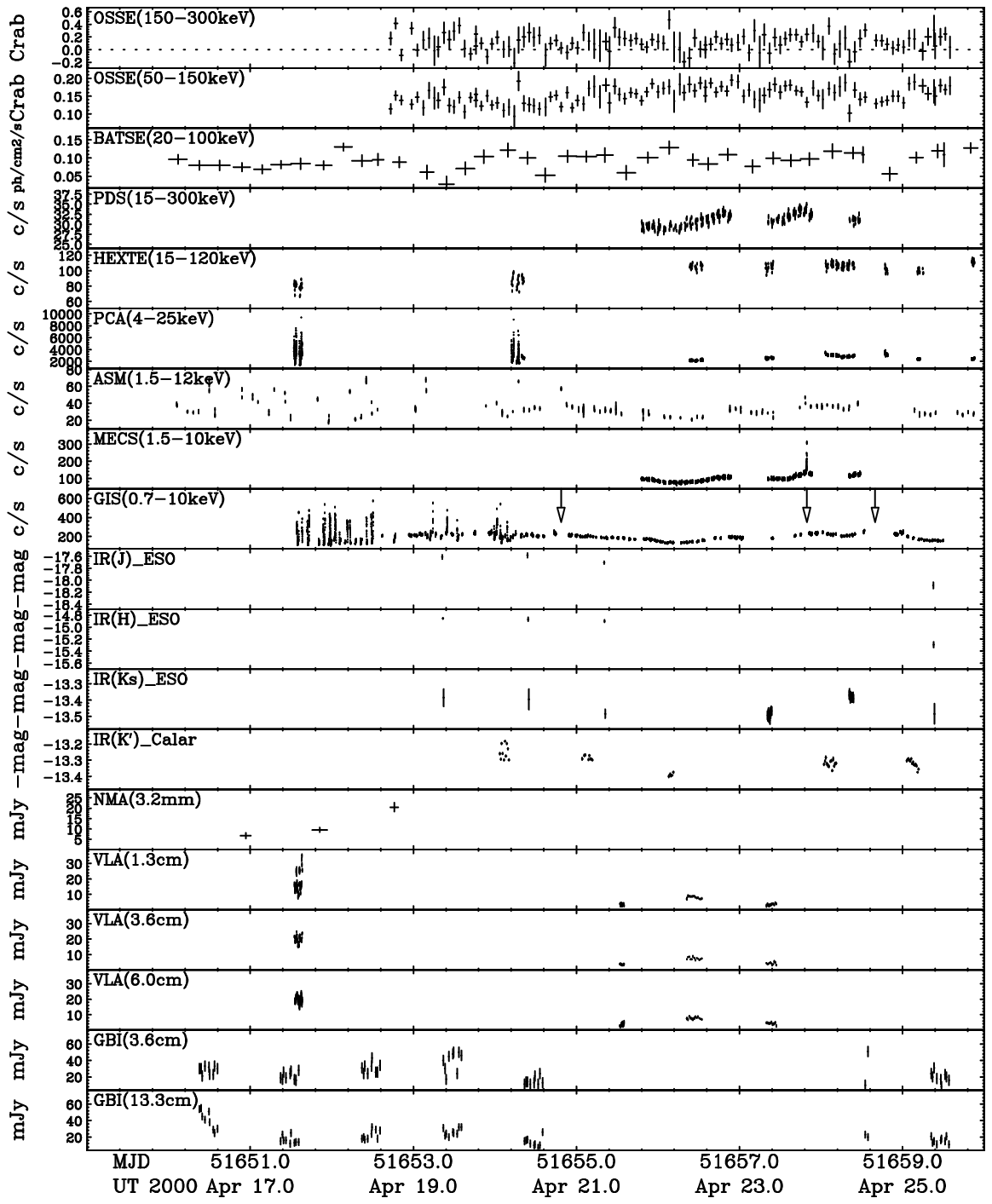]{The whole multiwavelength light curves of
GRS~1915+105 obtained from the 2000 April campaign, covering between
April 16 and 25. They are sorted in energy band (higher energy in
upper). From upper to lower panels, \cgro\ OSSE (150--300 keV: unit
Crab), OSSE (50--150 keV: Crab), \cgro\ BATSE (20--100 keV: photons
cm$^{-2}$ s$^{-1}$), \sax\ PDS (15--300 keV: counts s$^{-1}$), \rxte\
HEXTE (15--120 keV: counts s$^{-1}$, sum of cluster-0 and -1), \rxte\
PCA (4--25 keV: counts s$^{-1}$, sum of PCU0, 2 and 3; the data when
PCU3 was off are not plotted), \rxte\ ASM (1.5--12 keV: counts
s$^{-1}$), \sax\ MECS (1.5--10 keV: counts s$^{-1}$), \asca\ GIS
(0.7--10 keV: counts s$^{-1}$, sum of GIS2 and GIS3), near-infrared \J
- mag (ESO), \H -mag (ESO), \Ks - mag (ESO), \Kp -mag (Calar Alto),
NMA ($\lambda=$3.2 mm: mJy), VLA (1.3 cm: mJy), VLA (3.6 cm: mJy), VLA
(6.0 cm: mJy), GBI (3.6 cm: mJy), and GBI (13.3 cm: mJy). The arrows
in the GIS panel indicate the times of the first peak of soft X-ray
flares seen in the GIS medium bit-rate data.}

\figcaption[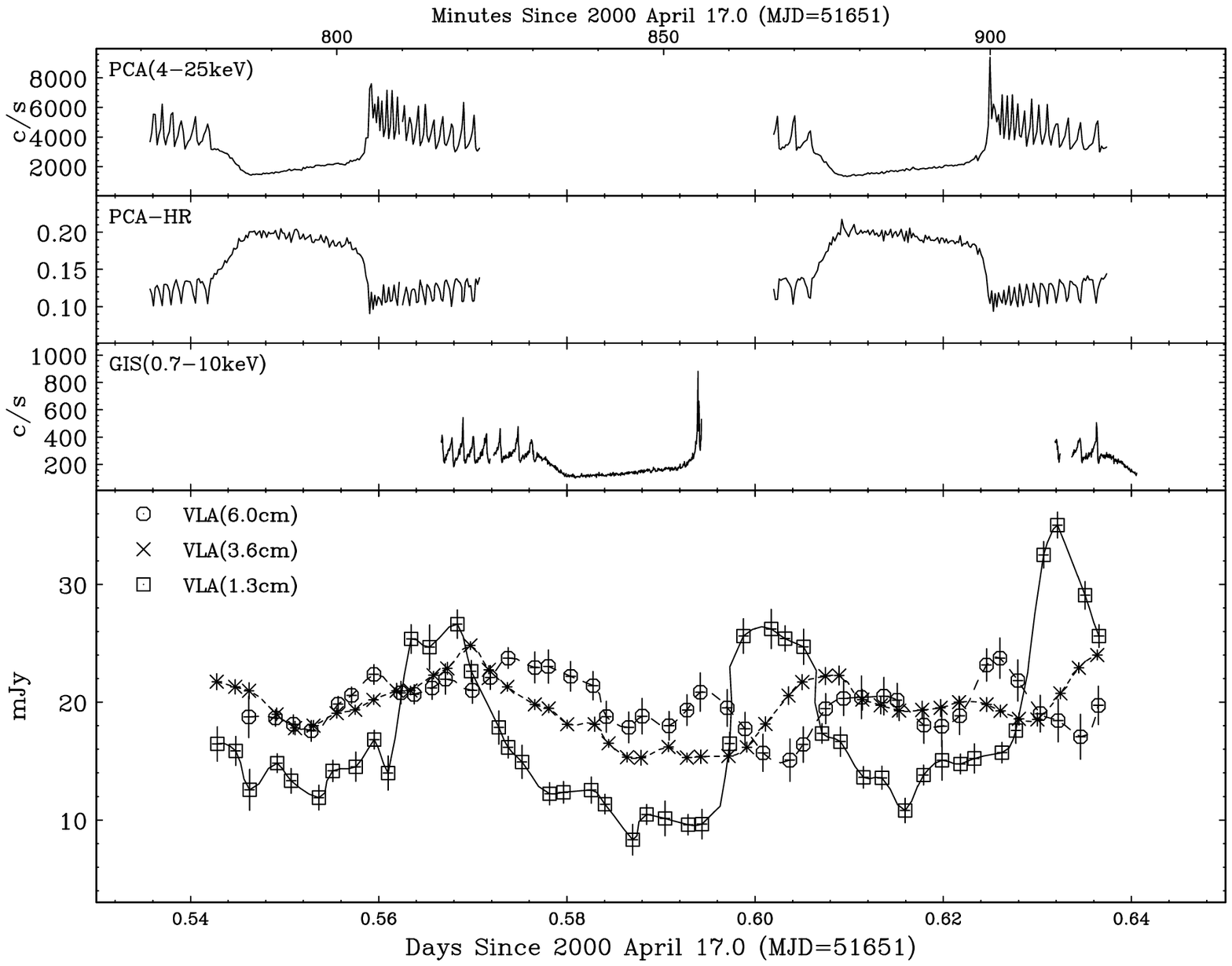]{
Multiwavelength light curves of \grs\ on 2000
April 17.53--17.65.  From upper to lower panels: \rxte\ PCA at 16 s
resolution (4--25 keV: unit counts s$^{-1}$), the PCA hardness ratio
4--10 keV and 10--25 keV, \asca\ GIS at 4 s resolution (0.7--10 keV:
counts s$^{-1}$), and radio fluxes by VLA (1.3 cm, 3.6 cm, and 6.0 cm:
mJy). }

\figcaption[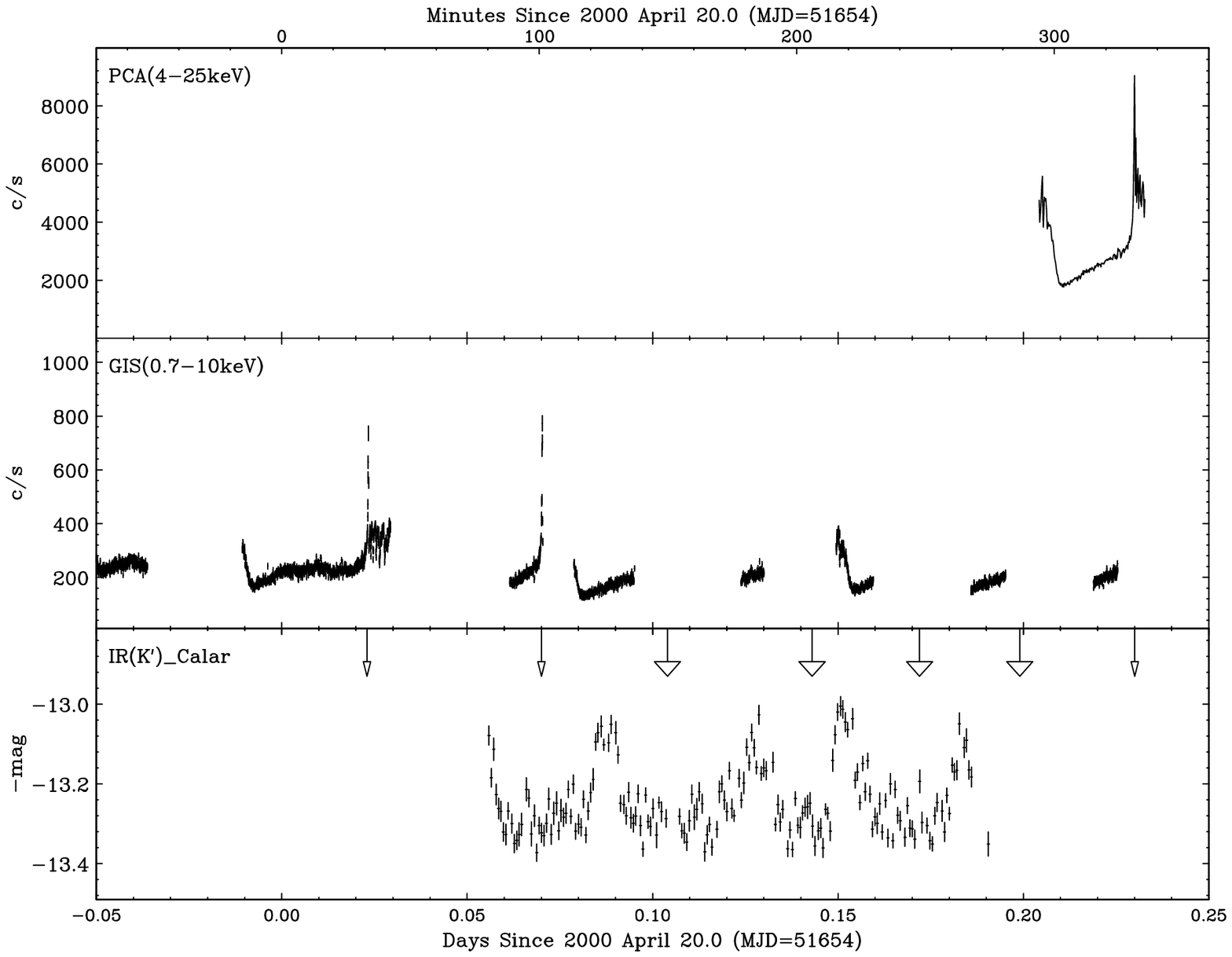]{
Multiwavelength light curves of \grs\ on 2000 April 19.95--20.25.
From upper to lower panels: \rxte\ PCA at 16 s resolution (4--25 keV:
unit counts s$^{-1}$), \asca\ GIS at 4 s resolution (0.7--10 keV: counts
s$^{-1}$), and infrared \Kp\ magnitude taken at the Calar Alto
observatory. The arrows with a large (small) width in the bottom panel
indicate the estimated (observed) time of the first peak of the
soft X-ray flares (see \S~3.3).
}

\figcaption[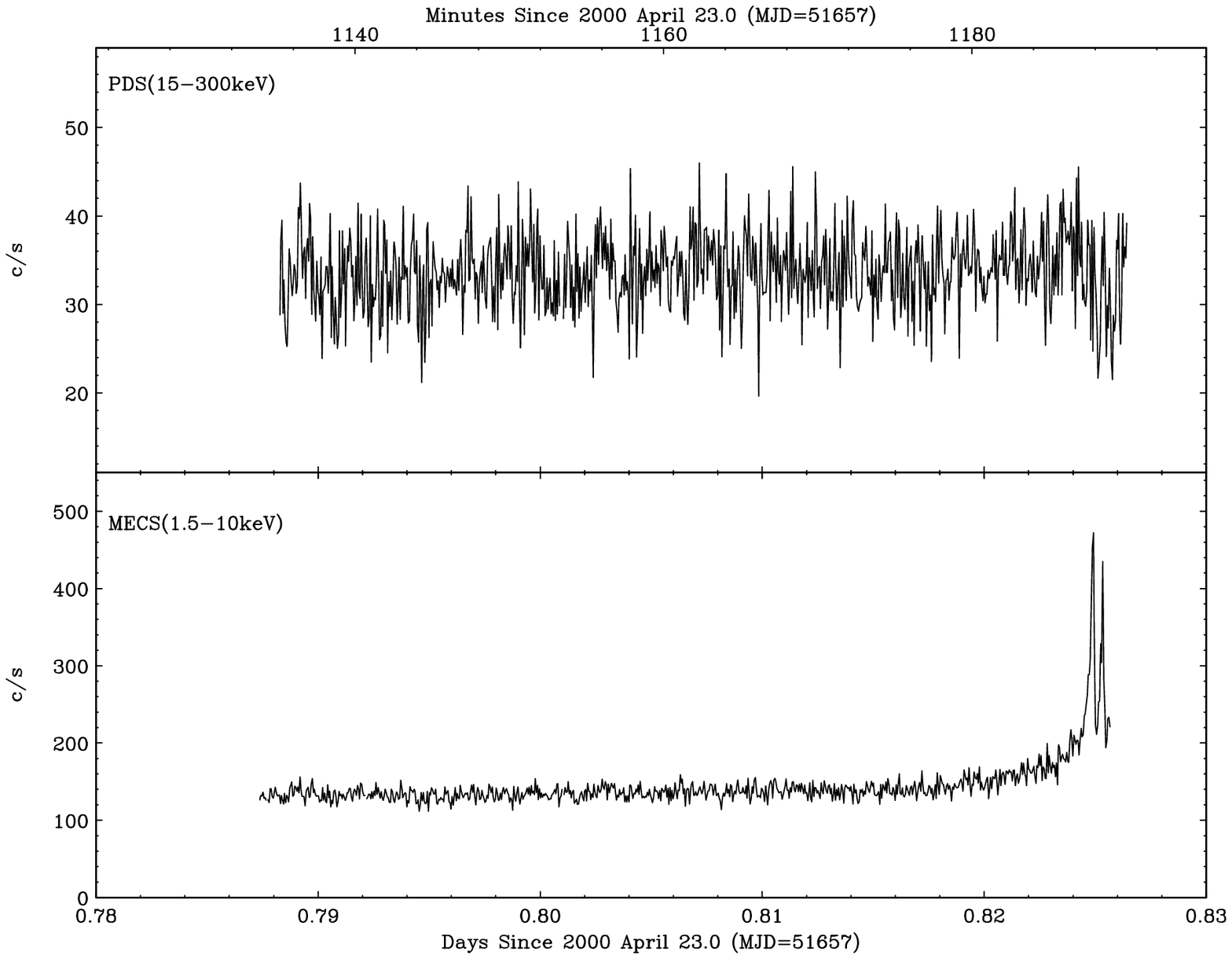]{
Multiwavelength light curves of \grs\ on 2000 April 23.78--23.83.
From upper to lower panels: \sax\ PDS at 4 s resolution (15--300 keV:
unit counts s$^{-1}$) and MECS (1.5--10 keV).
}

\figcaption[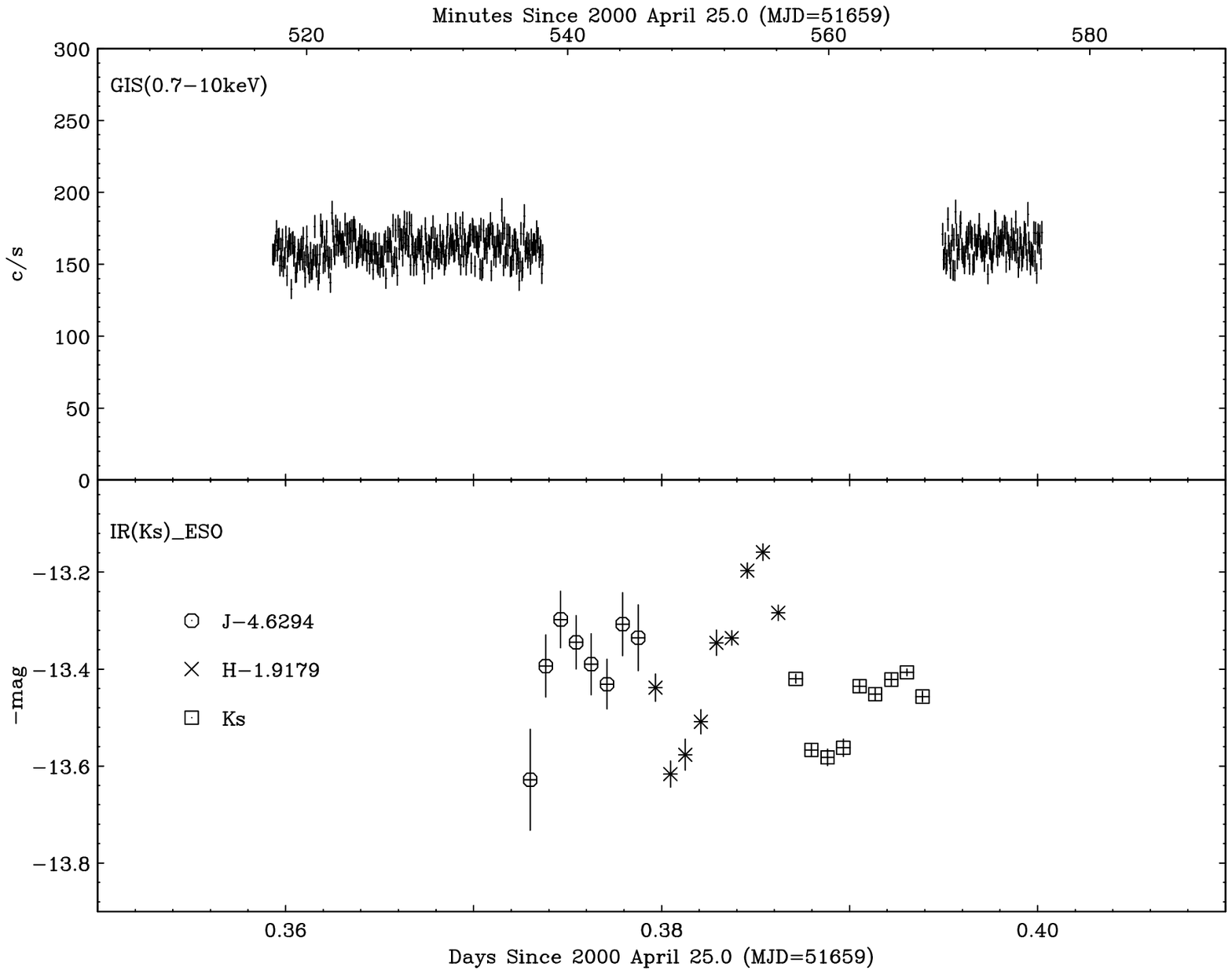]{
Multiwavelength light curves of \grs\ on 2000 April 25.35--25.41.
From upper to lower panels: \asca\ GIS at 4 s resolution (0.7--10 keV:
unit counts s$^{-1}$) and near-infrared magnitudes (\J , \H , and \Ks) taken 
at the ESO. Offsets of --4.6294 and --1.9179 are added to the \J\ and \H\
magnitudes, respectively. 
}

\figcaption[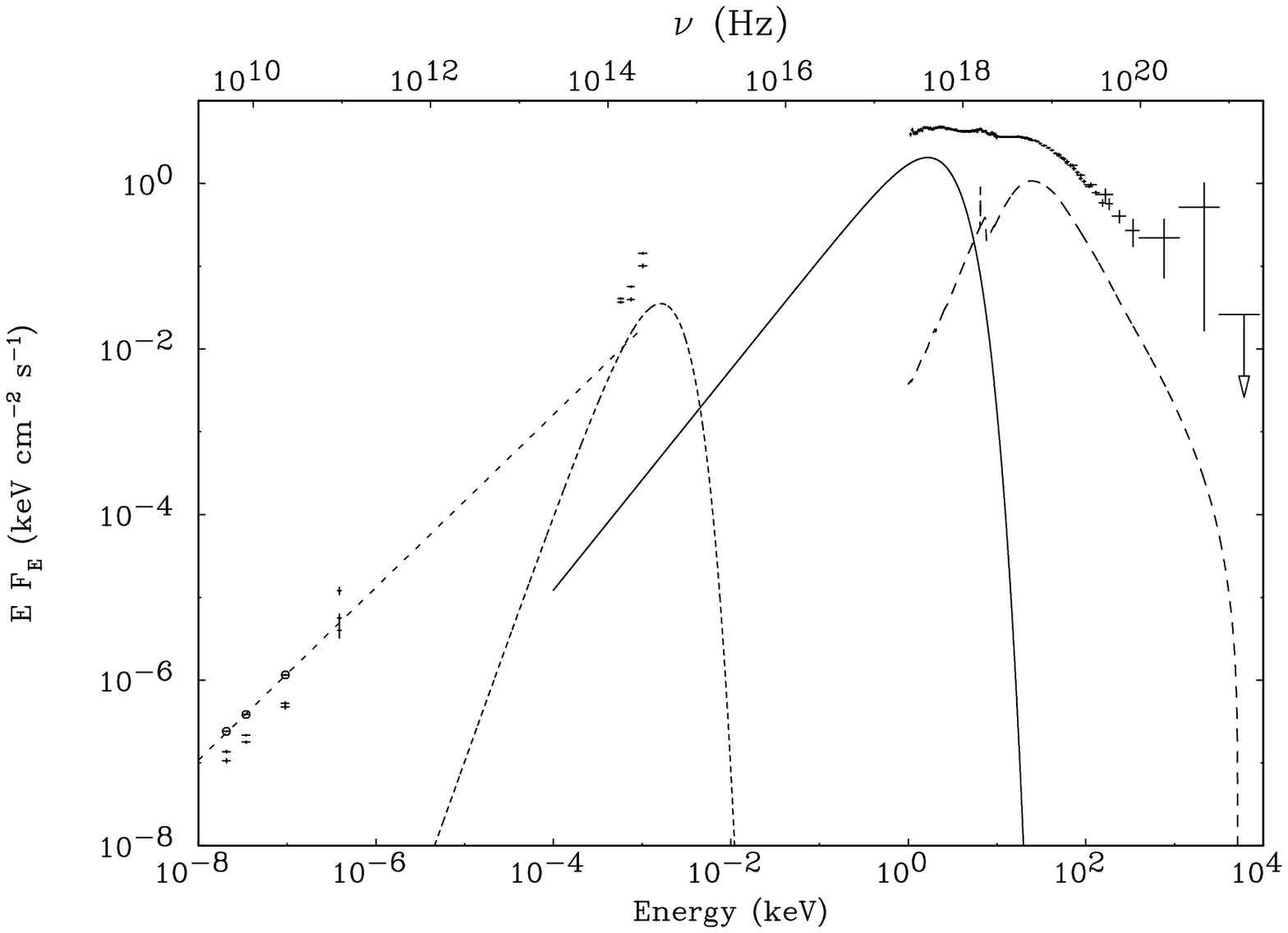]{ The spectral energy distribution of \grs\ in the
plateau state (except for the 94 GHz data) obtained from the campaign.
Interstellar absorption and extinction are corrected. The 1--200 keV
data are calculated from the simultaneous GIS-PCA-HEXTE spectra on
April 22.4--22.6 and the 0.05--10 MeV data are from the averaged OSSE
spectrum on April 21.0--26.0. The near-infrared data correspond to the
minimum and maximum flux densities during April 21.0--26.0 observed at
the ESO. The three fluxes in the radio band (5, 8, and 22 GHz) are
taken from the VLA data on April 21, 22, and 23, while those at 94 GHz
($\lambda$=3.2 mm) are from the NMA data on April 16, 17, and
18. Models from right to left: the long dashed line corresponds to an
estimated contribution of the reflection component ($\Omega / 2 \pi$ =
0.64, $\xi$ = 2; see Table~4 for detailed explanation) with an iron-K
emission line (at 6.5 keV with an equivalent width of 4 eV), the solid
line the multi-color disk ($kT_{\rm in}$ = 0.70 keV, $R_{\rm in}$ = 61
km), short-dashed line (right) the blackbody spectrum of the companion
star with a temperature of 4800 K, and the short-dashed straight line
(left) the best-fit power law determined from the 3 radio fluxes on
April 22 ($I_{\nu} \propto \nu^{0.04}$), which are simultaneous data
to the 1--200 keV spectra plotted here.}

\figcaption[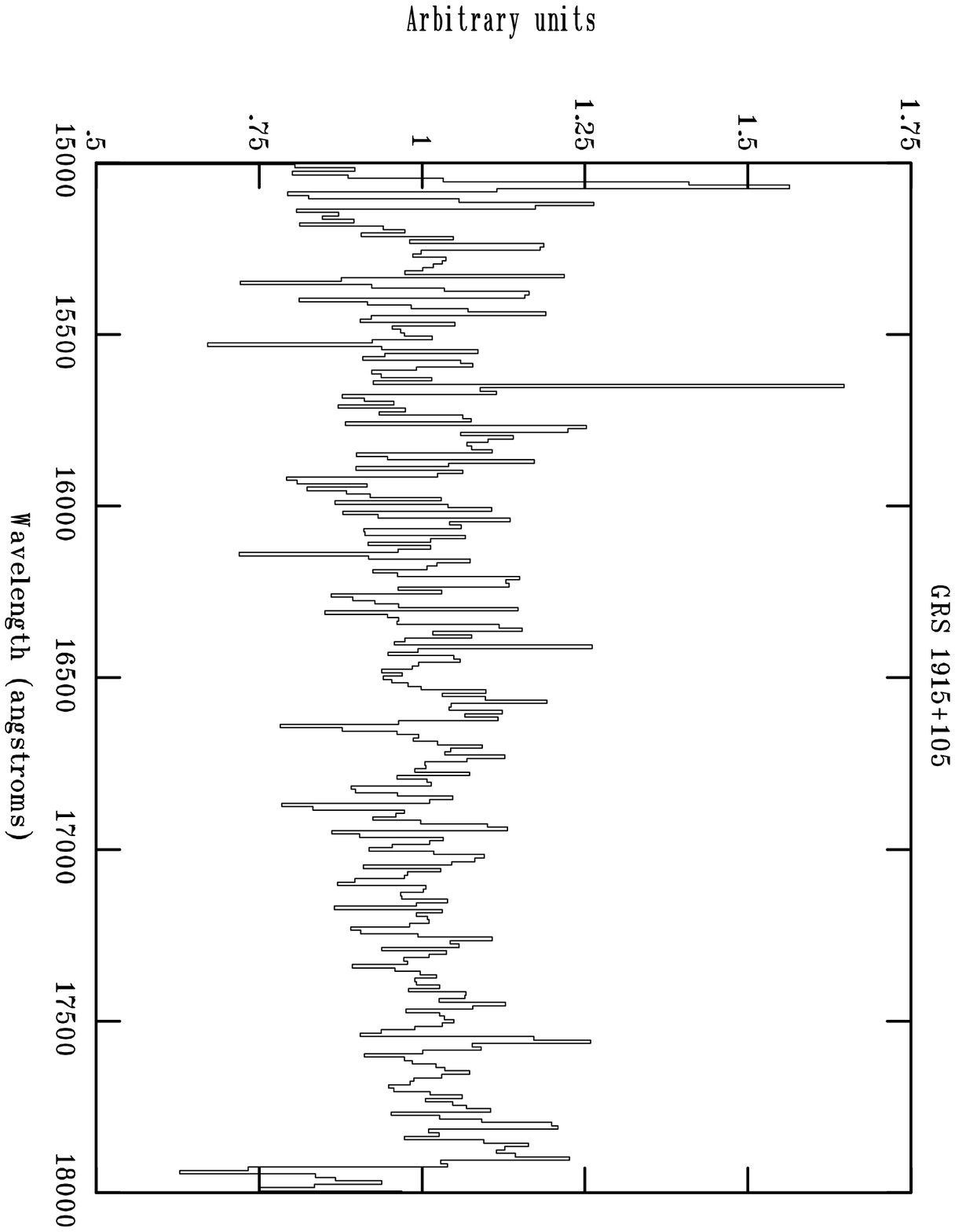,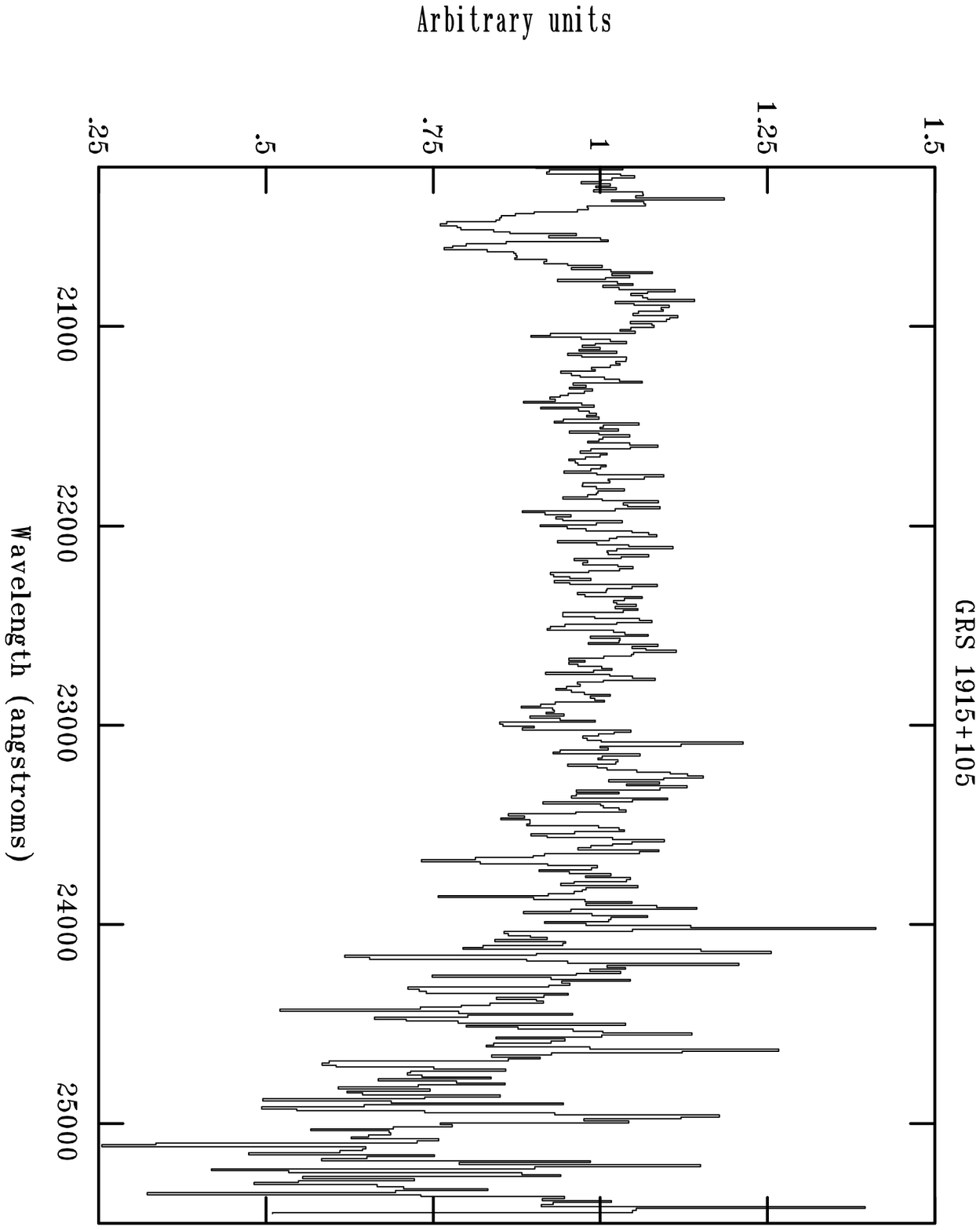]{
The infrared spectra of \grs\ taken at the ESO on 2000 April 22 at a
resolution of $\lambda/\Delta \lambda=$600 (upper: the 1.53--1.80
$\mu$m range, lower: 2.03--2.55 $\mu$m). The spectra are normalized 
and therefore vertical units are arbitrary.
}

\figcaption[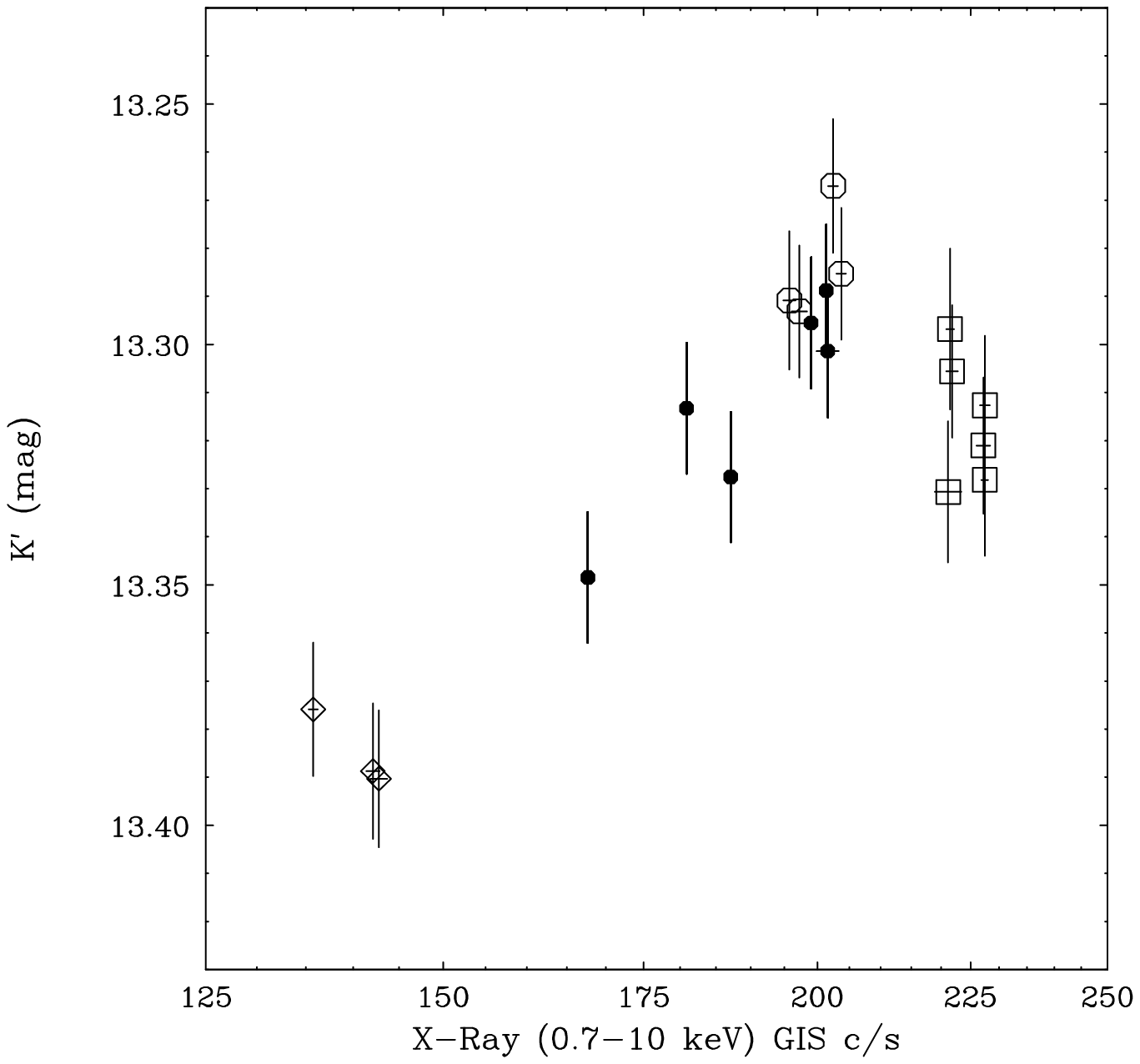]{
The correlation between the infrared \Kp\ magnitude (taken at Calar
Alto) and the GIS count rates (0.7--10 keV), each averaged for
$\sim$16 minutes with exactly simultaneous exposures. Different
symbols correspond to different days (open circles: April 21,
diamonds: April 22, open squares: April 24, filled circles: April 25).
The data of April 20 are not included, which showed infrared
flares. 
}

\figcaption[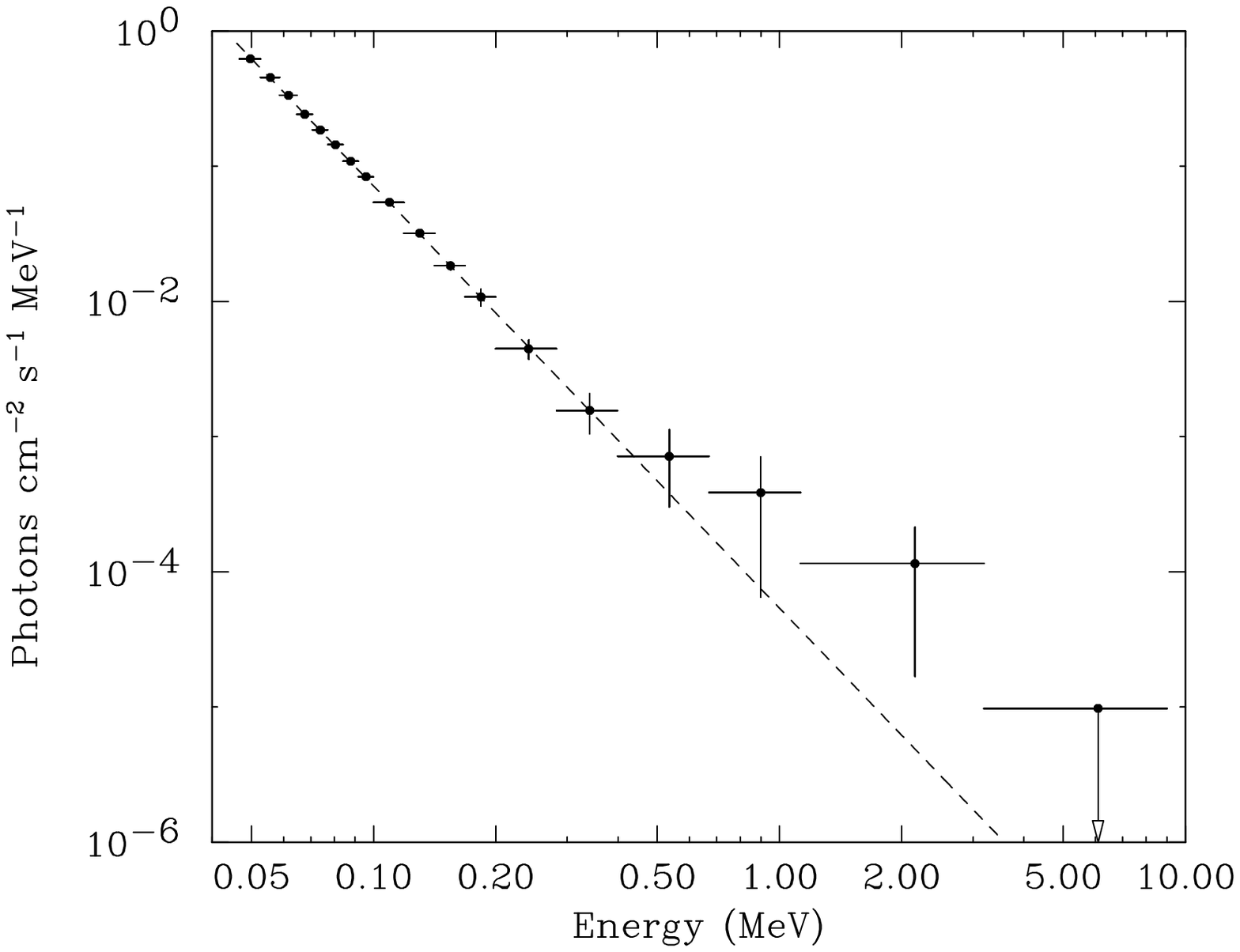]{
The OSSE unfolded spectrum in the 0.05--10 MeV range taken during 2000
April 18.7 to 25.6. The model is the best fit power law with a photon
index of 3.12. }

\figcaption[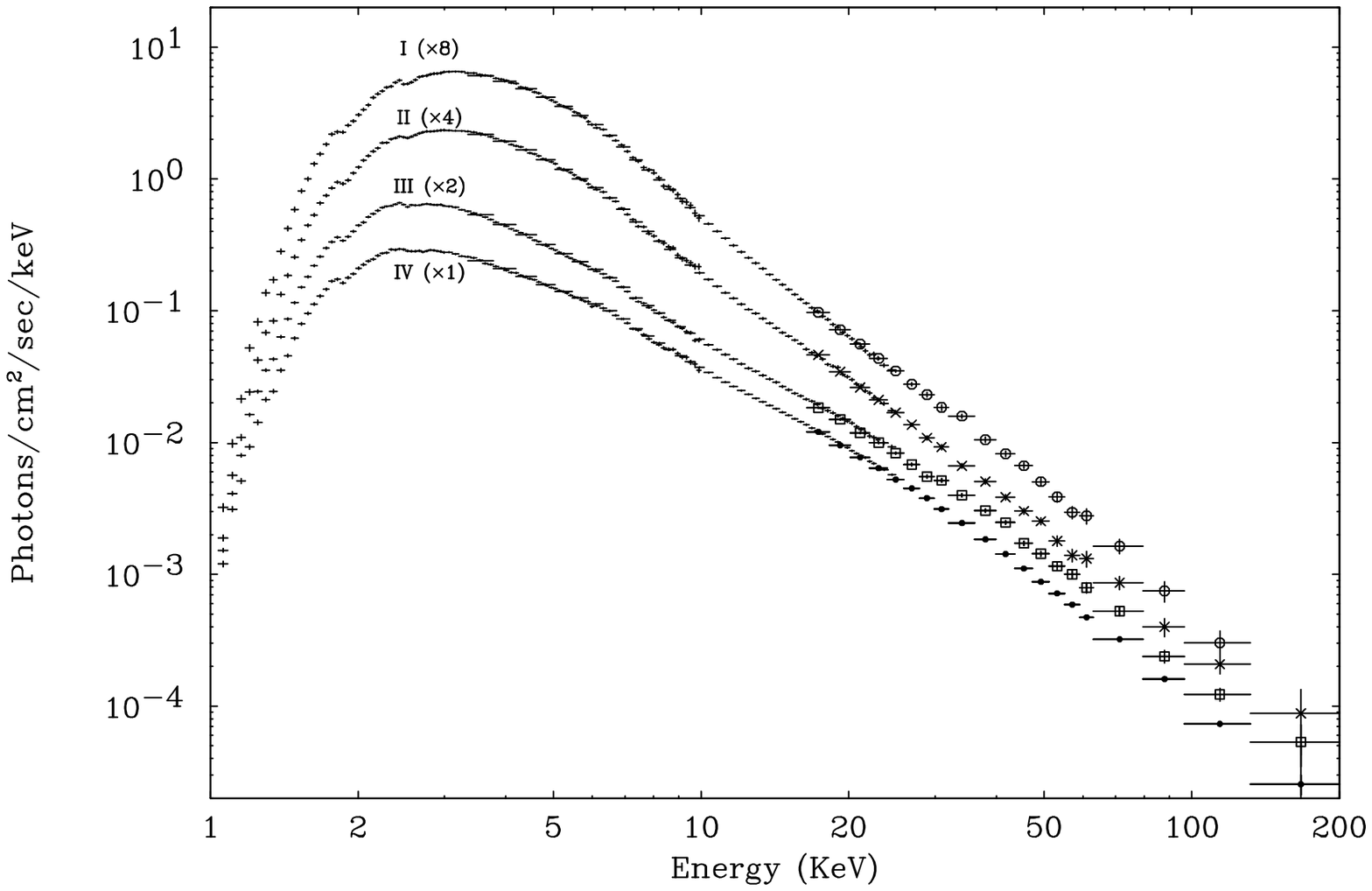]{
The combined GIS-PCA-HEXTE unfolded spectra in the 1--200 keV band 
in the four different states (see text), (I) the oscillation-high, 
(II) the oscillation-low, (III) the dip, and (IV) the plateau state, 
each multiplied by a factor of 8, 4, 2, and 1, respectively, for
clarity of the plots. 
}

\fi

\onecolumn
\begin{scriptsize}
\begin{deluxetable}{llll}
\tablenum{1}
\tablecaption{Log of the Multiwavelength Observations\label{tbl-1}}
\tablehead{\colhead{Observatory/Telescope} & \colhead{Instrument} &\colhead{Band} &\colhead{Observation Date (UT, 2000 Apr)} }
\startdata
{\it CGRO} & OSSE & 50 keV--10 MeV & 18.7--25.6 \nl
	   & BATSE & 20--100 keV    & monitor \nl
\tablevspace{0.3cm}
{\it BeppoSAX}  & PDS & 15--300 keV & 21.80--22.89, 23.35--23.89, 24.34--24.48 \nl
 \nodata	& MECS/LECS & 0.1--10 keV & same as above \nl
\tablevspace{0.3cm}
{\it RXTE} & HEXTE & 15--250 keV & 17.54--17.64, 20.20--20.36, 22.39--22.55, 23.32--23.42, 24.05--24.41, \nl
	&	&	& 24.78-24.82, 25.18--25.26, 25.85--25.88\nl
\nodata	& PCA & 2--60 keV & same as above \nl
\nodata	& ASM & 1.5--12 keV & monitor \nl
\tablevspace{0.3cm}
{\it ASCA} & GIS/SIS & 0.5--10 keV & 17.57--25.50 \nl
\tablevspace{0.5cm}
ESO / NTT 	& SOFI (Imaging) & $J$--$H$--$Ks$ & 19.35--19.37, 20.39--20.43, 21.34--21.36, 25.37--25.39 \nl
\nodata	& SOFI (Imaging) & $Ks$  & 23.34--23.40, 24.35--24.40 \nl
\nodata	& SOFI (Spectroscopy) &1.53--2.52 $\mu$m & 22.3 \nl
\tablevspace{0.3cm}
Calar Alto / 1.23m    & MAGIC	 & $K^{\prime}$ & 20.06--20.19, 21.07--21.21, 22.08--22.21, 24.03--24.20, 25.05--25.21 \nl
\tablevspace{0.3cm}
NMA	& 		& 88.6, 100.6 GHz & 16.88--17.01, 17.76--17.94, 18.72--18.81 \nl
\tablevspace{0.3cm}
VLA 	& & 5, 8, 22 GHz & 17.54--17.64, 21.53--21.58, 22.35--22.54, 23.33--23.45 \nl
\tablevspace{0.3cm}
GBI	& & 2.25, 8.3 GHz & monitor (except 21--24) \nl
\enddata
\end{deluxetable}
\end{scriptsize}

\clearpage
\begin{deluxetable}{lccc}
\tablenum{2}
\tablecaption{Averaged Near-Infrared Magnitudes of \grs\ Observed at the ESO \label{tbl-2}}
\tablehead{\colhead{Date (UT, 2000 Apr)} & \colhead{ \J } &\colhead{ \H } &\colhead{ \Ks } }
\startdata
19.35--19.38& 17.612$\pm$0.034 &14.856$\pm$0.012 &13.385$\pm$0.054 \nl
20.39--20.43& 17.586$\pm$0.037 &14.872$\pm$0.032 &13.396$\pm$0.063 \nl
21.33--21.36& 17.707$\pm$0.025 &14.900$\pm$0.022 &13.484$\pm$0.027 \nl
23.34--23.40& \nodata & \nodata & 13.490$\pm$0.032 \nl
24.34--24.40& \nodata & \nodata & 13.380$\pm$0.018 \nl
25.37--25.40& 18.086$\pm$0.054 & 15.290$\pm$0.048 &13.484$\pm$0.062 \nl
\enddata
\end{deluxetable}

%\clearpage
\begin{deluxetable}{lccc}
\tablenum{3}
\tablecaption{Line Features in the Infrared Spectra\label{tbl-3}}
\tablehead{\colhead{Id. ($\lambda$)} & \colhead{observed $\lambda$} &\colhead{E.W. ($\AA$)} &\colhead{FWHM ($\AA$)} }
\startdata
He I (2.0587) &  2.0558  & --8.83 & 37.3 \nl
Br gamma (2.1661) &  2.1600  & --2.16 & 19.14 \nl
He II (2.1891) & 2.1895  & --4.65  &39.25 \nl
Na I (2.20624--2.20897) & 2.2044  & --4.91 & 35.2 \nl
\enddata
\end{deluxetable}

\clearpage
\begin{deluxetable}{lccccc}
%\footnotesize
\scriptsize
\tablenum{4}
\tablecaption{Results of the Fit to the GIS-PCA-HEXTE Spectra\label{tbl-4}}
\tablehead{\colhead{Parameter} &\colhead{State~I} &\colhead{State~II} &\colhead{State~III}&\colhead{State~IV} &\colhead{State~IV} \nl
\colhead{} &\colhead{} &\colhead{} &\colhead{}&\colhead{(April 22)} &\colhead{(April 23)} \nl
}
\startdata
%\cutinhead{Continuum (GIS)}
$N_{\rm H}$ ($10^{22}$ cm$^{-2}$)
	&4.8$\pm$0.1 &4.6$\pm$0.1 &4.2$\pm$0.1 &4.0$\pm$0.1 &4.1$\pm$0.1 \nl
$kT_{\rm in}$ (keV)\tablenotemark{a} 
	&1.46$\pm$0.05 &1.28 $\pm$0.04 &0.85$\pm$0.02 &0.70$\pm$0.03 &0.84$\pm$0.05 \nl
$R_{\rm in}$ (km)\tablenotemark{a} 
	& 30$\pm$2 & 32$\pm$2 &52$\pm$3 &61$\pm$7 & 49$\pm$8 \nl
%$\Gamma_1$ & 1.65 (fix) & 1.65 (fix) & 1.65 (fix) & 1.65 (fix) & 1.65 (fix) \nl
$A$\tablenotemark{b}
	& 3.6$\pm$0.4& 2.8$\pm$0.2& 1.72$\pm$0.04 & 2.16$\pm$0.04& 2.44$\pm$0.09\nl
$\Gamma$\tablenotemark{b}
	& 2.60$\pm$0.10 & 2.44$\pm0.06$ & 2.52$\pm$0.05 &2.41$\pm$0.05 &2.54$\pm$0.09 \nl
$E_{\rm break}$ (keV)\tablenotemark{b}
	& 6.1$\pm$0.5 & 6.3$\pm$0.2 & 6.6$\pm$0.2 &6.5$\pm$0.2 &6.7$\pm$0.3 \nl
%$E_{\rm cut}$ (keV) & 17 (fix) & 17 (fix) & 17 (fix)& 17 (fix) & 17 (fix)\nl
$E_{\rm fold}$ (keV)\tablenotemark{b}
	& 115$^{+56}_{-32}$ & 78$^{+17}_{-12}$ & 1000 (fix) &200$^{+70}_{-40}$ & 290$^{+530}_{-120}$ \nl
\tablevspace{0.3cm} 
$\Omega / 2 \pi$\tablenotemark{c}& & & 0.88$\pm$0.10 &0.64$\pm$0.08 & 0.66$\pm$0.14 \nl
$\xi$\tablenotemark{c}	& & & $<$0.02 & $<$20 &$<$73 \nl
%Line Energy (keV)& & &6.5 (fix) &6.5 (fix) & 6.5 (fix) \nl 
E.W. (6.5 keV) (eV)\tablenotemark{d} & & & $<$12& $<$13& $<$21 \nl
%$E_{\rm edge}$ (keV)& & &9.3 (fix) &9.3 (fix) &9.3 (fix) \nl  
$\tau_{\rm edge}$ (9.3 keV)\tablenotemark{e} & & & 0.08$\pm$0.03 & 0.05$\pm$0.02 & 0.04$\pm$0.03 \nl 
\tablevspace{0.3cm}
2--10 keV Flux ($10^{-8}$)\tablenotemark{f}
	&2.23& 1.54& 0.77& 0.80& 0.97 \nl
10--200 keV Flux ($10^{-8}$)\tablenotemark{f}
	&0.98& 0.88& 0.88& 1.09& 1.10 \nl
$\chi^2$/dof  &174/154 &178/154& 163/151 & 142/150 &103/150 \nl
\tablenotetext{a}{ 
The component of the multi-color disk model (Mitsuda \etal\ 1984):
$kT_{\rm in}$ and $R_{\rm in}$ is the (apparent) innermost temperature
and radius, respectively. To calculate $R_{\rm in}$
a distance of 11 kpc and an inclination of 66$^{\circ}$ are assumed.
}
\tablenotetext{b}{ 
The component of a broken power law: $A$ is the normalization at 1 keV
in units of photons cm$^{-2}$ s$^{-1}$ keV$^{-1}$, $\Gamma$ is the
photon index above the break energy $E_{\rm break}$, below which the
photon index is fixed at 1.65. High energy cutoff (``highecut'' in XSPEC) is
multiplied with an cutoff energy of 17 keV (fixed) with a folding
energy $E_{\rm fold}$.}
\tablenotetext{c}{ 
The reflection component by Magdziarz \& Zdziarski (1995, ``pexriv''
in XSPEC). $\Omega$ is the solid angle of the reflector and $\xi$ is the ionization parameter. The disk temperature, elemental abundance, and inclination
are fixed at $3\times10^{4}$K, 1 solar, and 66$^\circ$,
respectively. The input spectrum is approximated by a cutoff power law
model with a photon index and a folding energy of 2.4 and 200 keV
(State~III), 2.0 and 70 keV (State~IV, April 22), and 2.2 and 80 keV
(State~IV, April 23).}
\tablenotetext{d}{ 
The equivalent width of an iron-K emission line fixed at 6.5 keV with a 1$\sigma$ width of 10 eV.}
\tablenotetext{e}{ 
The optical depth of the 9.3 keV edge from H-like iron ions.
An absorption line from the same ions are included 
with the parameters as measured by \chandra\ HETGS (Lee \etal\ 2002).
}
\tablenotetext{f}{ 
The absolute flux is normalized to the GIS value.
}
\tablecomments{Errors are 90\% confidence limits for a single parameter.}
\enddata
\end{deluxetable}

% Figures
\clearpage
\noindent
Figure 1

\vspace{1cm}
\noindent
\begin{center}
\mbox{\psfig{figure=f1.ps,width=0.90\textwidth,angle=0}}
\end{center}

\clearpage
\noindent
Figure 2

\noindent
\begin{center}
\mbox{\psfig{figure=f2.ps,height=0.4\textheight,angle=0}}
\end{center}

\vspace{1.5cm}

\noindent
Figure 3

\noindent
\begin{center}
\mbox{\psfig{figure=f3.ps,height=0.4\textheight,angle=0}}
\end{center}

\clearpage
\noindent
Figure 4

\noindent
\begin{center}
\mbox{\psfig{figure=f4.ps,height=0.4\textheight,angle=0}}
\end{center}

\vspace{1.5cm}

\noindent
Figure 5

\noindent
\begin{center}
\mbox{\psfig{figure=f5.ps,height=0.4\textheight,angle=0}}
\end{center}

\clearpage
\noindent
Figure 6
\vspace{1cm}

\noindent
\begin{center}
\mbox{\psfig{figure=f6.ps,width=\textwidth,angle=0}}
\end{center}

\clearpage
\noindent
Figure 7

\noindent
\begin{center}
\mbox{\psfig{figure=f7a.ps,width=0.7\textwidth,angle=90}}
\mbox{\psfig{figure=f7b.ps,width=0.7\textwidth,angle=90}}
\end{center}

\clearpage
\noindent
Figure 8

\noindent
\begin{center}
\mbox{\psfig{figure=f8.ps,height=0.4\textheight,angle=0}}
\end{center}

\vspace{1.5cm}

\noindent
Figure 9

\noindent
\begin{center}
\mbox{\psfig{figure=f9.ps,height=0.4\textheight,angle=0}}
\end{center}

\clearpage
\noindent
Figure 10
\vspace{1cm}

\noindent
\begin{center}
\mbox{\psfig{figure=f10.ps,width=\textwidth,angle=0}}
\end{center}

\end{document}